\documentclass[twocolumn]{aastex631}

\usepackage{pifont}

\begin{document}

\title{Assessing robustness and bias in 1D retrievals of 3D Global Circulation Models\\at high spectral resolution: a WASP-76 b simulation case study in emission}
\shorttitle{Robustness of 1D Retrievals with 3D HRS}

\author[0000-0002-9610-3367]{Lennart van Sluijs}
\affiliation{Department of Astronomy, University of Michigan, 1085 S. University Ave., Ann Arbor, 48109, MI, USA}
\correspondingauthor{Lennart van Sluijs}
\email{lsluijs@umich.edu}

\author[0000-0002-6980-052X]{Hayley Beltz}
\affiliation{Astronomy Department, University of Maryland, College Park, 4296 Stadium Dr., College Park, MD 207842 USA}

\author[0000-0003-0217-3880]{Isaac Malsky}
\affil{Jet Propulsion Laboratory, California Institute of Technology, Pasadena, CA 91109, USA}

\author[0009-0004-0227-4205]{Genevieve H. Pereira}
\affiliation{Department of Astronomy, University of Michigan, 1085 S. University Ave., Ann Arbor, 48109, MI, USA}

\author[0000-0002-6311-4860]{L. Cinque}
\affiliation{Department of Astronomy, University of Michigan, 1085 S. University Ave., Ann Arbor, 48109, MI, USA}

\author[0000-0003-3963-9672]{Emily Rauscher}
\affiliation{Department of Astronomy, University of Michigan, 1085 S. University Ave., Ann Arbor, 48109, MI, USA}

\author[0000-0002-4125-0140]{Jayne Birkby}
\affiliation{Astrophysics, University of Oxford, Denys Wilkinson Building, Keble Road, Oxford, OX1 3RH, UK}

\begin{abstract}
High-resolution spectroscopy (HRS) of exoplanet atmospheres has successfully detected many chemical species and is quickly moving toward detailed characterization of the chemical abundances and dynamics. HRS is highly sensitive to the line shape and position, thus, it can detect three-dimensional (3D) effects such as winds, rotation, and spatial variation of atmospheric conditions. At the same time, retrieval frameworks are increasingly deployed to constrain chemical abundances, pressure-temperature (P-T) structures, orbital parameters, and rotational broadening. To explore the multidimensional parameter space, they need computationally fast models that are consequently mostly one-dimensional (1D). However, this approach risks introducing interpretation bias since the planet's true nature is 3D. We investigate the robustness of this methodology at high spectral resolution by running 1D retrievals on simulated observations in emission within an observational framework using 3D Global Circulation Models of the quintessential HJ WASP-76 b. We find that the retrieval broadly recovers conditions present in the atmosphere, but that the retrieved P-T and chemical profiles are not a homogeneous average of all spatial and phase-dependent information. Instead, they are most sensitive to spatial regions with large thermal gradients, which do not necessarily coincide with the strongest emitting regions. Our results further suggest that the choice of parameterization for the P-T and chemical profiles, as well as Doppler offsets among opacity sources, impact retrieval results. These factors should be carefully considered in future retrieval analyses.
\end{abstract}


\section{Introduction} \label{sec:intro}
\begin{deluxetable*}{l c c r}
\tablecaption{An overview of relevant WASP-76 system parameters and reported uncertainties. This is to show WASP-76 b is a quintessential HJ. \label{table:w76_system_params}}
\tablewidth{0pt}
\tablehead{
\colhead{\textbf{Quantity}} & \colhead{\textbf{Symbol}} & \colhead{\textbf{Value}} & \colhead{\textbf{Reference}}
}
\startdata
\quad Stellar radius & $R_{\star}$ & $1.73 \pm 0.04 \ \text{R}_{\sun}$ & \citet{West2016} \\ 
\quad Stellar effective temperature & $T_{\star}$ & $6250 \pm 100 \ \text{K}$ & \citet{West2016} \\ 
\hline
\quad Primary transit time & $T_{0}$ & 2456107.85507 $\pm$
0.00034 (JD) & \citet{West2016} \\ 
\quad Eccentricity & $e$ & 0\tablenotemark{1} & \citet{West2016} \\ 
\quad Inclination & $i$ & $88 \pm 1.6 \ \rm{deg}$ & \citet{West2016} \\ 
\quad Orbital period & $P$ & $1.809886\pm0.000001 \ \rm{d}$ & \citet{West2016} \\ 
\quad Planetary radius & $R_{\text{p}}$ & $1.83 \pm 0.06 \ \rm{R}_{\rm{Jup}}$ & \citet{West2016} \\ 
\quad Gravitational acceleration & $\log{g}$ & 2.80 $\pm$ 0.02 (cgs) & \citet{West2016} \\
\quad Effective temperature & $T_{\rm{p}}$ & $2160 \pm 40 \ \rm{K}$ & \citet{West2016} \\ 
\hline
\quad System velocity & $V_{\text{sys}}$ & $-7.0 \pm 0.1 \ \rm{km/s}$ & \citet{Pelletier2023} \\ 
\quad Keplerian velocity & $K_{\text{p}}$ & $180.7 \pm 0.6 \ \rm{km/s}$ & \citet{Pelletier2023} \\
\quad Rotational velocity & $v_{\text{rot}}$ & 5.3 km/s\tablenotemark{2} & This work \\
\enddata
\tablenotetext{1}{Adopted by \citet{West2016}, who constrained $e<0.05 \ \rm{at} \ 3\sigma$.}
\tablenotetext{2}{Computed from the planet's period and radius assuming a synchronized orbit and solid body rotation.}
\end{deluxetable*}
Exoplanet atmospheric characterization of hot Jupiters (HJs, $T_{\rm{eq}}>1000$ K), is quickly transitioning from detecting atmospheric species to detailed characterization of atmospheric chemistry and dynamics \citep{Fortney2021}. One major goal is to link the current atmospheric composition to formation pathways, through observable metrics such as the C/O-ratio, metallicity, and refractory-to-volatile ratios \citep[e.g.][]{Oberg2011, Lothringer2021, Khorsid2022, Chachan2023}. Another key goal is understanding their atmospheric physics under extreme conditions not found in the solar system.
High-resolution spectroscopy (HRS, R $\geq$ 15,000, see \citet{vanSluijs2023}) has successfully constrained HJ chemistry and dynamics for over a decade \citep[for a recent review see][]{Snellen2025}. This method uses the large orbital acceleration of the planet relative to the star during an observing night to disentangle an exoplanet atmosphere from stellar and telluric signatures. The earliest detections targeted strong atomic lines such as the sodium doublet \citep[e.g][]{Narita2005, Redfield2008, Snellen2008, Wyttenbach2015}. HRS has also been combined with cross-correlation spectroscopy to detect atmospheric species with faint, but numerous spectral lines \citep[e.g.][]{Snellen2010, Birkby2018, Snellen2025}. This is done by employing cross-correlation with a spectral template to combine the signal of many spectral lines with an individual signal-to-noise ratio (S/N) $S/N_{\text{line}}<1$. There are many detected species\footnote{See this Github repository aiming to contain an overview of all HRS literature: \url{https://github.com/arjunsavel/hires-literature/tree/main?tab=readme-ov-file}} including a range of ions and atoms \citep[e.g.][]{Hoeijmakers2020b, Wardenier2021, Kesseli2022, Bello-Arufe2022, Prinoth2022, Silva2022, Borsato2023, Gandhi2023} and molecules \citep[e.g.][]{Snellen2010, Brogi2012, Birkby2013, Giacobbe2021, vanSluijs2023, Pelletier2023}.
One major advantage of HRS is its ability to detect the shapes and shifts of the planet's spectral features as these features are spectroscopically resolved. This makes HRS particularly sensitive to Doppler broadening caused by the combined effects of day-to-nightside winds, jets, and planetary rotation. Frequently, measured orbital velocities of individual species are offset from expectations \citep[e.g.][]{Snellen2010, Brogi2016, Wardenier2021, Kesseli2021, Gandhi2022, Pino2022, Brogi2023, Nortmann2024, Simonnin2024, Seidel2025} which have been attributed to day-to-nightside winds and eastward jets. Doppler broadening of spectral lines has also been observed \citep[e.g.][]{Snellen2014, Brogi2016}. Some studies investigated the phase-dependent variation of the spectral line strength, which can be attributed to offsets of the hottest point on these HJs away from the sub-stellar point \citep{Kawahara2012, Herman2022, Hoeijmakers2022, Pino2022, vanSluijs2023,  Beltz2024, Lesjak2024}. \citet{Kesseli2024} pioneered a new way to use weak and strong iron lines to probe different atmospheric pressure regions and measure vertical wind shear.
Recently, HRS atmospheric retrieval frameworks have been developed \citep{Brogi2019, Gibson2022} and are used for detailed atmospheric constraints. To explore the multidimensional parameter space, these atmospheric retrievals link a sampling algorithm such as {\sc Pymultinest} \citep{Feroz2008, Feroz2009, Feroz2019} or {\sc dynesty} \citep{Speagle2020}, to a forward model that quickly computes and compares spectra to observations. They have been used to constrain chemical and pressure-temperature (P-T) profiles, abundance ratios, metalicity, rotational broadening, eccentricity, and orbital phase-offsets \citep[e.g.][]{Brogi2019, Line2021, Gibson2022, Mansfield2024, Smith2024, Finnerty2024, Kanumalla2024, Bazinet2024, Blain2024, Cont2024, Debras2024}.
Models of different dimensionality have been employed to do HRS. One-dimensional (1D) models, such as {\sc PHOENIX} \citep{Hauschildt1997}, {\sc petitRADTRANS} \citep{Molliere2019}, or {\sc POSEIDON} \citep{MacDonald2017, MacDonald2023, Wang2025}, calculate the exoplanet spectrum from a single P-T structure and chemical profiles as a function of atmospheric pressure, frequently using parameterized functions \citep[e.g.][]{Madhusudhan2009, Guillot2010}. 1D models have the advantage over three-dimensional (3D) models by being computationally less expensive but at the cost of simplifying the complex 3D nature of a real exoplanet atmosphere. 3D Global Circulation Models (GCMs) include longitudinal and latitudinal variations of the P-T and chemical profiles including the effect of global wind patterns and rotation. They are computed by solving the fluid dynamical equations globally coupled with a treatment for radiative heating and cooling \citep[e.g.][]{Showman2020}. Several GCMs have been developed to simulate HJ atmospheric conditions \citep[e.g.][]{Showman2002, Cho2003, DobbsDixon2008, Showman2009, Rauscher2010, Heng2011, Mayne2014, Mendonca2016}. GCMs have been used to predict observable trends in both high-resolution emission and transmission spectroscopy \citep{Ricci2012, Showman2013, Zhang2017, Malsky2021, Harada2021, Beltz2022, Lee2022, Beltz2023, Wardenier2023, Savel2023, Beltz2024, Wardenier2025} and have been used to model HRS data to good effect \citep[e.g.][]{Flowers2019, Beltz2021, vanSluijs2023}.
Despite the observational evidence and theoretical predictions indicating the 3D nature of HJ atmospheres, HRS retrievals have been limited to using 1D forward models, due to the required large number of model evaluations \mbox{($\sim 10^{5}-10^{6}$)}. \citet{Beltz2022} found a higher detection S/N when using a 3D template over a grid of hundreds of 1D models in their HRS analysis of HD 209458 b. Although they assessed the impact of model dimensionality on the measured detection significance, no study has investigated the robustness and bias of using 1D forward models within retrievals on inherently 3D HRS data.
This work explores the intricate nuances of using 1D retrievals on 3D HRS observations. This is achieved by running retrievals on simulated HRS observations using phase-dependent spectra computed from a GCM within a simple observational framework. We choose to include only the minimal observational effects necessary to run a 1D HRS retrieval, to focus on the impact of 3D effects. For a recent study investigating how observational effects impact retrievals, we refer the reader to \citet{Savel2024}. This study analyzes a GCM of WASP-76 b for hypothetical IGRINS \citep{Park2014, Levine2018} dayside post-eclipse observations. The choice for IGRINS was motivated by the extensive prior research with this instrument using HRS retrievals to obtain detailed atmospheric thermal and chemical profiles \citep[e.g.][]{Line2021, Brogi2023, Smith2024, Mansfield2024, Kanumalla2024, Smith2024b, Pelletier2024}. Consequently, investigating 3D bias and robustness affecting such analyses is beneficial. WASP-76 b is a quintessential HJ around a bright F star ($\sim$11.7 in V-band) with an equilibrium temperature of \mbox{$\sim2200 \ \text{K}$} \citep{West2016}. System parameters are summarized in Table~\ref{table:w76_system_params}. This exoplanet has been surveyed using a range of HRS instruments using both emission and transmission spectroscopy, which confirmed the presence of many atomic, ionic, and molecular species \citep[][]{Seidel2019, Ehrenreich2020, Tabernero2021, Kesseli2022, Sanchez-Lopez2022, Silva2022, Yan2023, Gandhi2023, CostaSilva2024, Mansfield2024, Wardenier2023, Lampon2023}. Although this work uses simulations of WASP-76 b and IGRINS, we emphasize that the primary goal is not to characterize this exoplanet or instrument in detail but rather to understand the nuances of running 1D retrievals on an inherently 3D dataset and its impact on atmospheric constraints and interpretation.
In Section~\ref{sec:3d_gcm}, we describe the analysis and results of the GCM of WASP-76 b to produce phase-dependent high-resolution emission spectra. In Section~\ref{sec:sims} we describe how these spectra are used to simulate HRS data within an observational framework. In Section~\ref{sec:retrieval} we describe our setup and results for running the 1D retrieval on the 3D simulated data. Section~\ref{sec:discussion} discusses the retrieval constraints in the context of our GCM. Finally, Section~\ref{sec:conclusion} summarizes and concludes this work and we highlight suggestions for future work.
\section{3D GCM: Analysis and Results} \label{sec:3d_gcm}
\subsection{Model setup and analysis}
\begin{figure*}
    \centering
\includegraphics[width=0.64\textwidth]{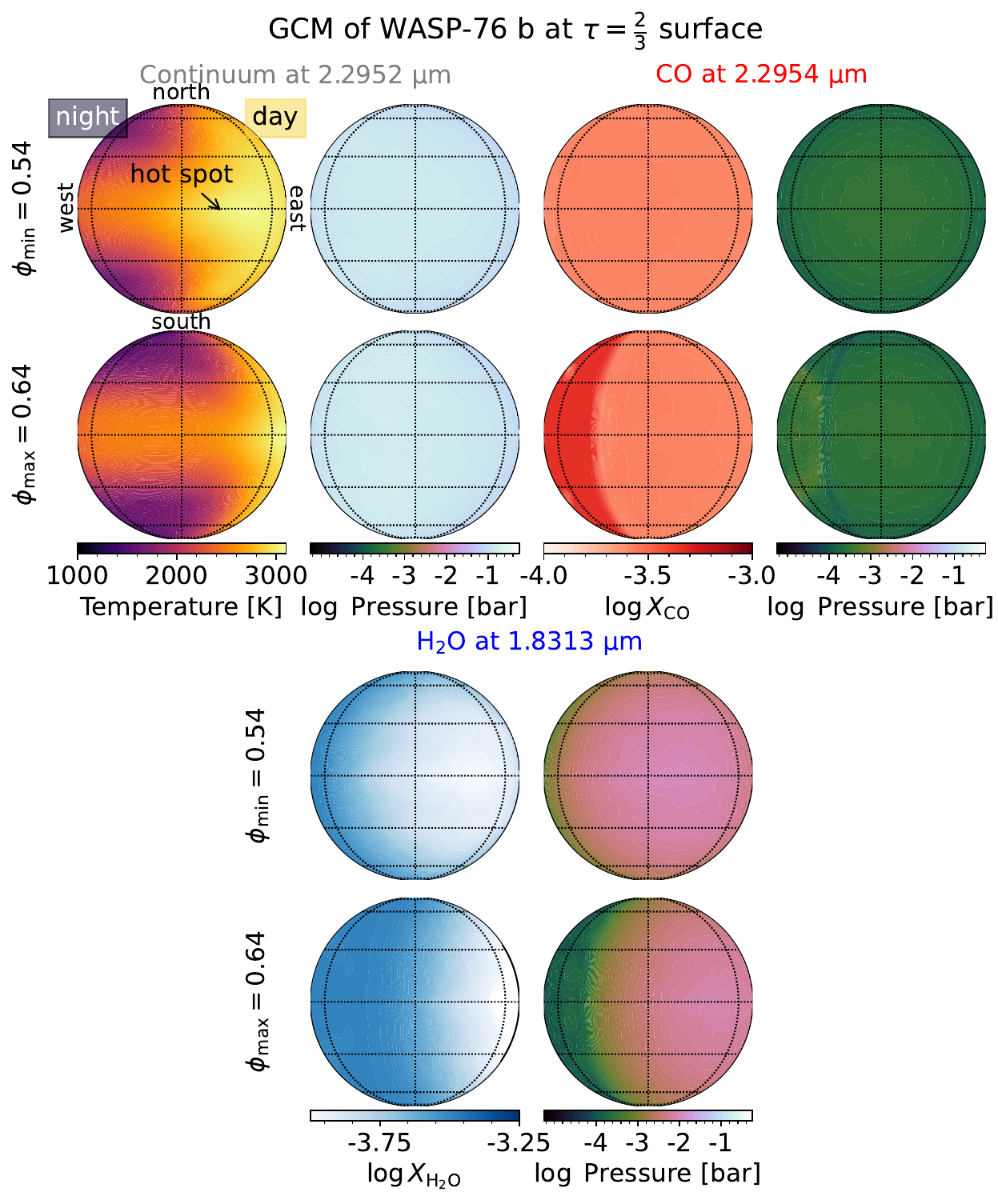} 
    \caption{Maps of WASP-76 b computed from the GCM output to demonstrate the variation of line-forming pressure and chemical abundances as a function of orbital phase and wavelength. Two phases are shown: at the start ($\phi_{\rm{min}}=0.54$) and at the end ($\phi_{\rm{max}}=0.64$) of the simulated HRS observations. The maps are shown for three wavelengths coinciding with the spectral continuum, the core of a CO spectral line and the core of a H$_2$O spectral line. The $\tau=\frac{2}{3}$ surface corresponds to the pressure level where these spectral features form at each wavelength. Each grid displays latitudes of 0$\degr$, $\pm30\degr$, and $\pm60\degr$, as well as the longitude closest to the observer at each phase, $lon(\phi)$, and at $lon(\phi) \pm 60\degr$. \textit{First column}: the= temperature at the continuum level. An eastward-offset hot spot is close to half phase, and a cooler nightside becomes visible towards quadrature phases. \textit{Second, fourth, and sixth columns:} the pressure where the line cores form for the continuum, H$_2$O, and CO. The continuum level forms around 100 mbar. On the dayside, CO lines form at lower pressures ($\sim$0.1 mbar) than H$_2$O lines ($\sim$1 mbar). On the nightside, they form around the same pressures ($\sim$1 mbar) due to increased water abundance. \textit{Third and fifth columns:} the chemical abundances for CO and H$_2$O, respectively, at the pressures where their line cores form. The effect of water dissociation can be seen around the hot spot.}
    \label{fig:gcm_maps} 
\end{figure*}
We use the 3D GCM RM-GCM\footnote{The code is publicly available here: \url{https://github.com/emily-rauscher/RM-GCM}} to model the planet WASP-76 b. These models appeared in \citet{Beltz2022} and \citet{Beltz2024}, but we summarize the key points here. The GCM solves the primitive meteorology equations; a standard, simplified set of the equations of fluid dynamics \citep{Rauscher2010}. These equations are coupled with a radiative transfer scheme which sets the heating or cooling rates in different wavelength bands. The radiative transfer scheme has two modes, which are "double-gray" \citep[two wavelength bands, see][]{Beltz2022} or "picket fence" \citep[five wavelength bands, see][]{Malsky2024}, and these modes are compared in \citet{Beltz2024}. We use the picket fence scheme, which is slightly more complex and should be physically more accurate at minimal additional computational expense \citep{Lee2021}. The GCM has three drag modes: drag-free, uniform drag, or active drag; where atmospheric drag adds additional resistance against the gas flows. In the active drag model, the local drag timescale depends on the local temperature, density, and chosen global surface strength. For our setup of WASP-76 b, we use the drag-free model, which produces a stronger eastward hot spot offset. The GCM outputs the temperature and wind structures throughout the atmosphere, in latitude, longitude, and pressure (the vertical coordinate). Chemical abundances are calculated assuming local chemical equilibrium and Solar elemental abundances.
To calculate simulated spectra from the 3D models, we run a line-by-line post-processing routine \citep[e.g. for validation of this routine, see][]{Beltz2022, Malsky2024, Beltz2024} through the atmosphere, using ray tracing to capture the correct geometry. These spectra are phase-dependent as they include the geometrical orientation of the exoplanet's hemispheric disk as it orbits its host star. The calculation can be done in two modes: Doppler-off or Doppler-on. The Doppler-on mode shifts the local opacities by the line-of-sight velocity from winds and rotation \citep{Zhang2017}, while the Doppler-off does not. To model the exoplanet's 3D structure as realistically as possible, the Doppler-on mode is used.
As motivated in the introduction, we simulate high-resolution observations for IGRINS. In its wavelength range from 1.4-2.6 $\mu$m CO and H$_2$O spectral features are expected to be prevalent, assuming equilibrium chemistry. They are the two main carbon- and oxygen-bearing species, key in constraining the C/O-ratio. Additionally, OH and H$^-$ are anticipated to contribute to opacity. For the most realistic spectrum to compare against real observations, OH and H$^-$, and additional trace species should be considered. However, as we will demonstrate, the interplay between thermal structure and chemical profiles is complex, even when focusing on just two species. As our goal is to understand this complexity, we have included only CO and H$_2$O. These were included in our radiative transfer calculation using the {\sc ExoMol} line lists for CO \citep{Li2015} and H$_2$O, respectively  \citep{Furtenbacher2020a, Furtenbacher2020b}.
To understand at which pressures the spectral lines form, we computed where the optical depth $\tau=2/3$ following \citet{Zhang2017}. In short, this is done during the post-processing of the GCM on an altitude grid with 250 layers. Including all opacity sources, the post-processing code is used to compute the optical depth at a specific wavelength by integrating along the observer's line of sight until $\tau=2/3$. This computation was done at three wavelengths: 2.2952 $\mu$m for the continuum,  2.2954 $\mu$m for a CO line core, and 1.8313 $\mu$m for a H$_2$O line core.
We also aim to understand how different regions on the planet disk contribute towards the total spectrum.  The total planet spectrum will be the summed intensity of each line-of-sight integrated over the full planet disk. Line-of-sights intersecting hotter layers of the atmosphere will contribute more flux than cooler regions, if seen at the same angle. The integration also accounts for spherical geometry, where regions closer to the edge of the disk will contribute less to the total spectrum, as they have a smaller angular extent as seen from the observer's point of view. This is done by weighting the intensity from each longitude ($lon$) and latitude ($lat$) by a geometric factor $f$
\begin{equation}
    {f = \cos^2{(lon)} * \cos(lat) \times dlon \times dlat},
\end{equation}
where the longitude and latitude are normalized such that the sub-observer point corresponds with $(lon, lat) = (0\degr, 0\degr)$. To understand the interplay between photons and geometry, we modified the post-processing to compute local spectra at each longitude and latitude before summing their contributions. To avoid additional complexity, this calculation was done in Doppler-off mode, as the primary goal is to understand the link between regional thermal structure and the global planet spectrum.
\subsection{GCM results}
A key result from the $\tau=2/3$ calculation is that the continuum, H$_2$O, and CO spectral features form at different atmospheric pressure levels. This is summarized in Figure~\ref{fig:gcm_maps}, which shows maps of the temperature, chemical abundance, and pressure levels where spectral lines form at two post-eclipse orbital phases (these phases are chosen at the start and the end of our simulated HRS observations as will be discussed in more detail in Section~\ref{sec:sims}). The temperature map shows an eastward hot spot offset directly after the secondary eclipse ($\phi=0.54$). Towards quadrature phases ($\phi=0.64$), the hot spot rotates to the edge of the disk and more of the cooler nightside becomes visible. Both CO and H$_2$O have a higher nightside chemical abundance at $\tau=2/3$, but the abundance increase of CO is smaller than that of water. The slight nightside increase in CO is attributed to small changes in pressure and temperature at the $\tau=2/3$ surface compared to the dayside. A depletion of the water chemical abundance is seen in the hot spot region due to water dissociation into OH and H$^-$. For water lines, the increase of water abundance on the night side makes line cores form higher up in the atmosphere. In summary, we find the spectral continuum level forms $\sim$100 mbar, the CO line core on both the day and nightside form $\sim$0.1 mbar, and the H$_2$O line cores form between $\sim$0.1-10 mbar.
\begin{figure*}
    \centering
    \includegraphics[width=0.45\textwidth]{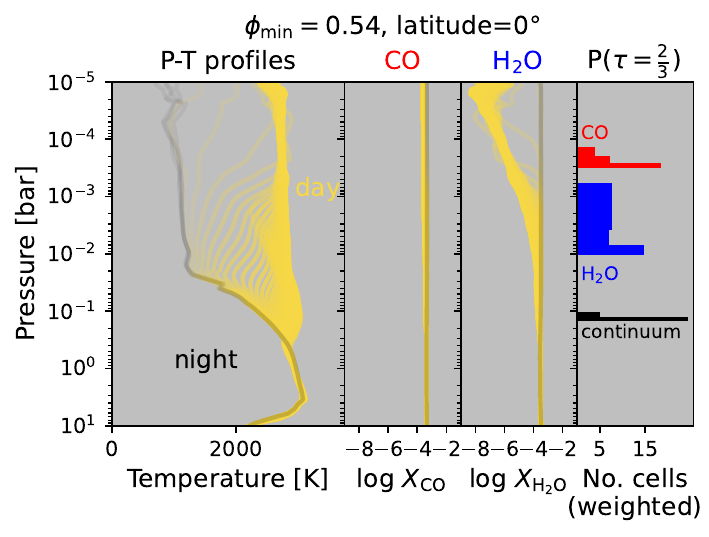}
    \includegraphics[width=0.45\textwidth]{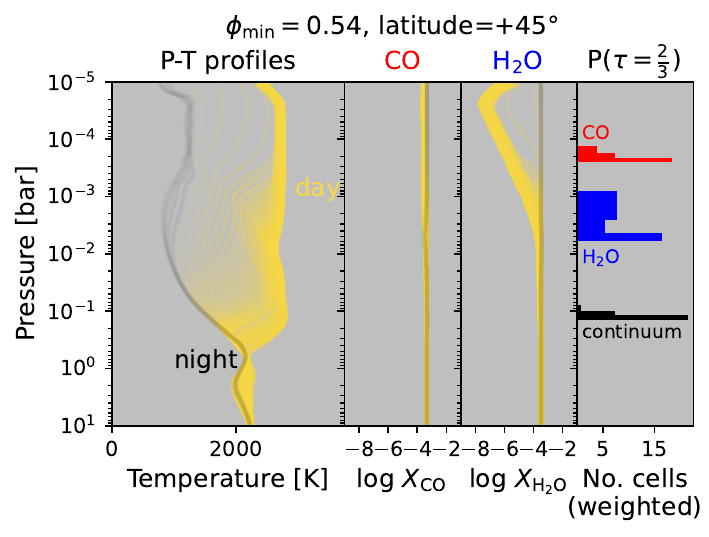}
    \includegraphics[width=0.45\textwidth]{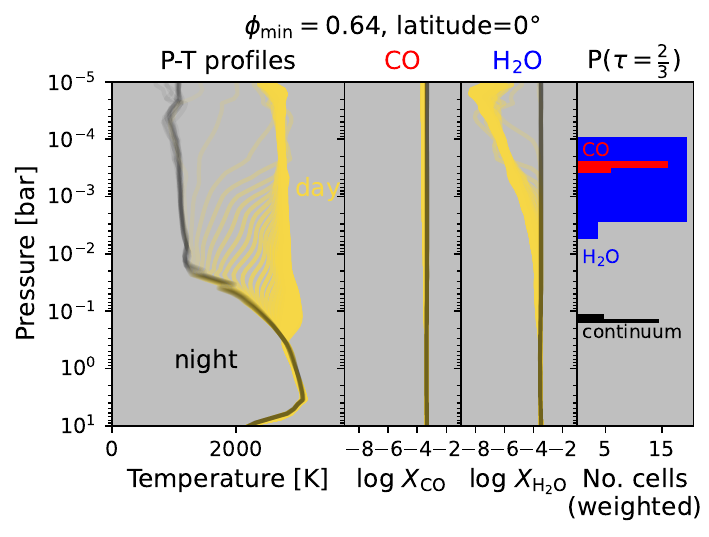}
    \includegraphics[width=0.45\textwidth]{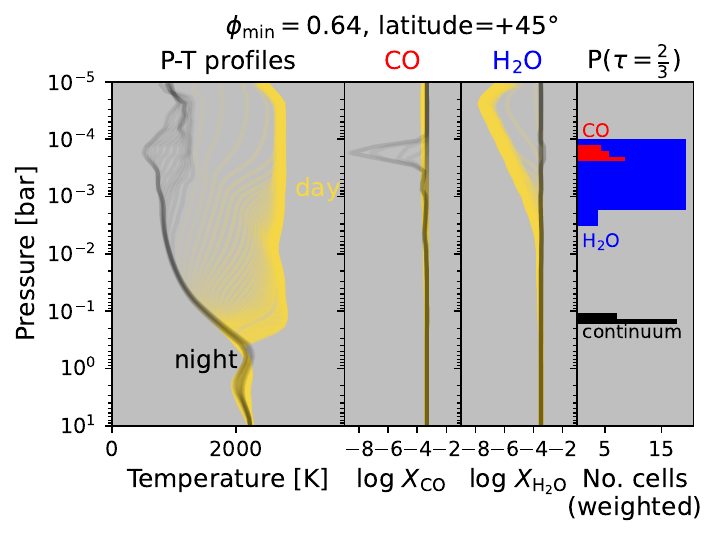}
    \caption{P-T and chemical profiles to show variation as a function of spatial location and observed phase. Profiles are shown at a latitude of $0\degr$ and $+45^\circ$ for longitudes visible at the start and end orbital phase. To visualize the effect of the geometry, the opacity of each longitudinal profile has been weighted by its geometric factor. The left panel shows the range of visible P-T profiles, where dayside profiles are colored yellow and nightside profiles are black. The middle panels show chemical profiles for CO and H$_2$O. CO has a fairly constant dayside chemical abundance, but water dissociation can be seen in the H$_2$O dayside profiles. CO abundances decrease only for the coldest nightside profiles. The rightmost panel shows a histogram counting the number of atmospheric cells, weighted by the geometric factor, for pressures where CO and H$_2$O line cores and the spectral continuum form. They are colored red for CO, blue for H$_2$O, and black for the continuum. The shape and strength of CO and H$_2$O lines are sensitive to pressures between the line core and spectral continuum forming regions, thus covering both inverted and non-inverted parts of the P-T profiles.}
    \label{fig:gcm_profiles} 
\end{figure*}
Figure~\ref{fig:gcm_profiles} visualizes the P-T and chemical profiles. The hottest dayside profiles show nearly isothermal P-T structures, whereas the coolest dayside profiles show strong inversion in the upper atmosphere and non-inversion in the deeper atmosphere. The CO profiles show fairly constant chemical abundance for the visible parts of the day and nightside, except for the coldest profiles reaching temperatures $<1000 \ \rm{K}$ (bottom-right panel of Fig.~\ref{fig:gcm_profiles}), where the CO abundance drops as methane (CH$_4$) is preferentially formed. The effect of water dissociation is seen in the upper atmospheres of the dayside P-T profiles.

\begin{figure}[h!]
    \centering
    \includegraphics[width=0.45\textwidth]{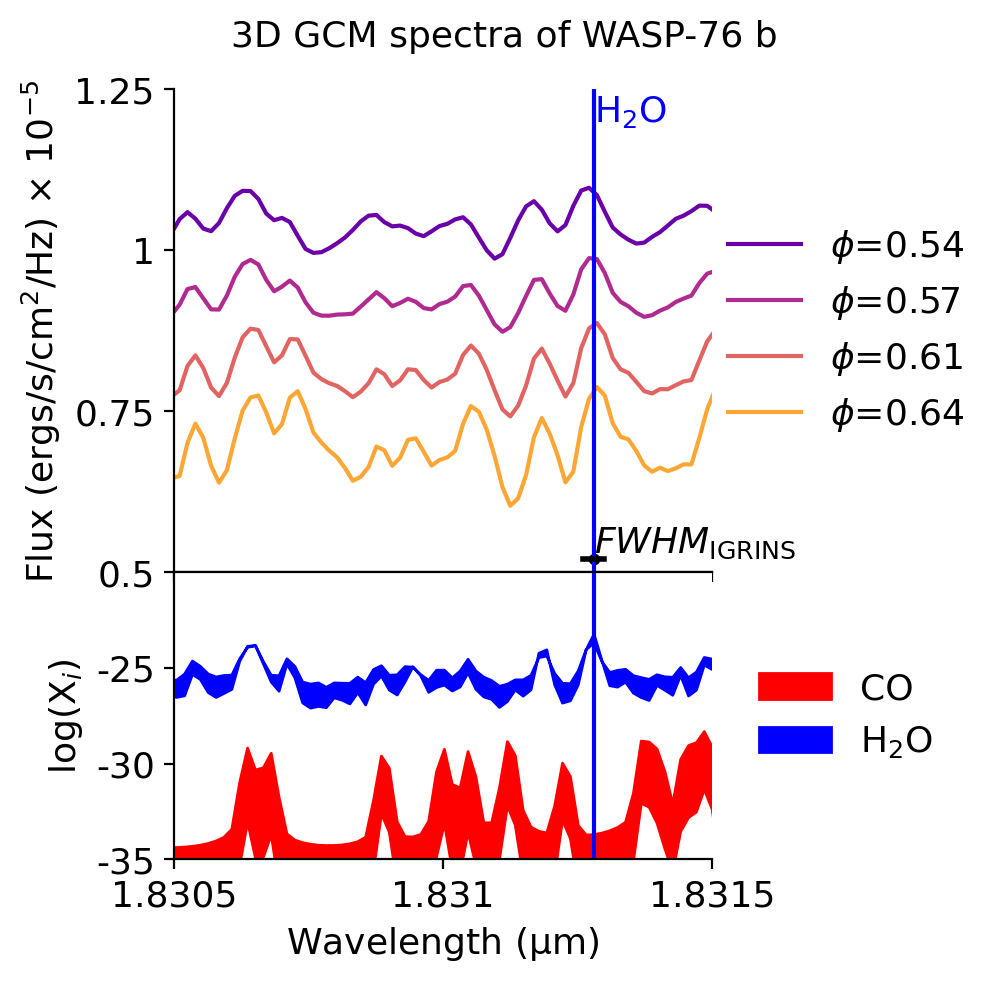}
    \includegraphics[width=0.45\textwidth]{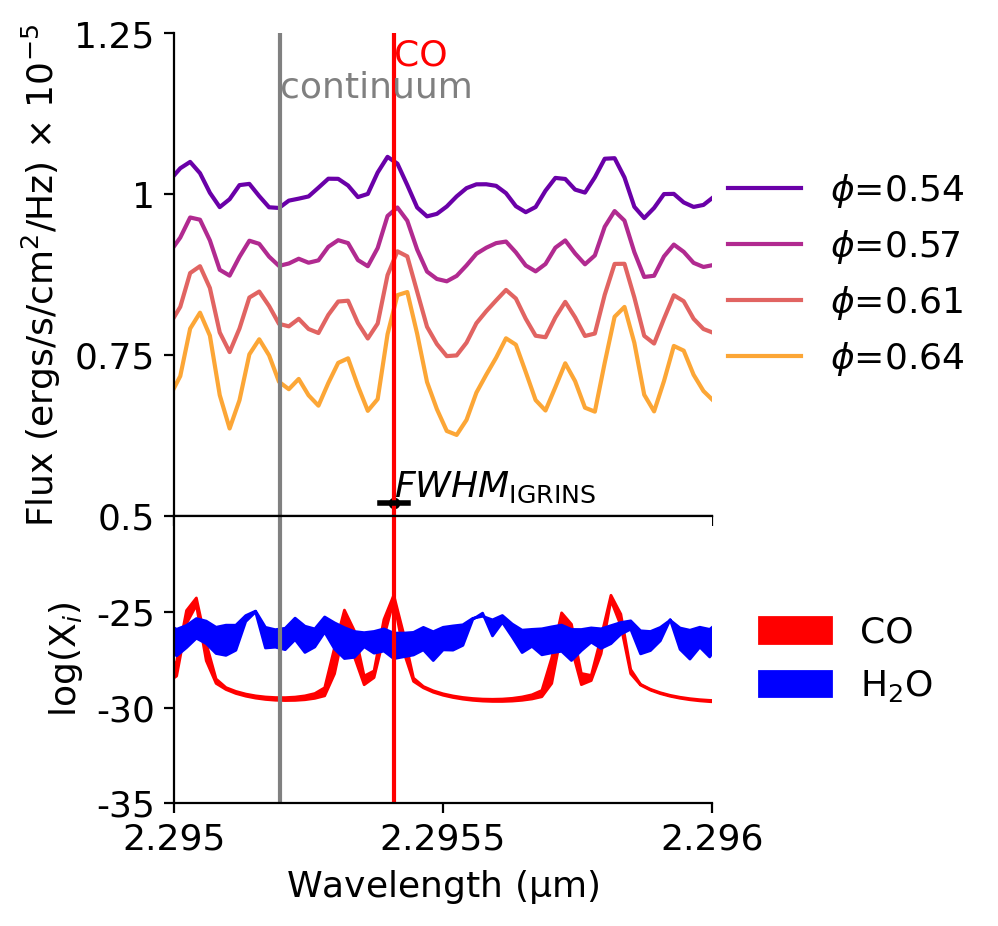}
    \caption{Phase-dependent GCM spectra show the complexity that arises from their underlying 3D P-T, chemical and wind profiles. For wavelengths around strong H$_2$O lines (left) and strong CO lines (right). We also show the cross-section ($\log X_i$) for H$_2$O and CO for the temperature range of 1500-2500 K at a pressure of 1 mbar. The vertical lines indicate wavelengths at a H$_2$O spectral line (blue), CO spectral line (red), and at the spectral continuum (gray). The black bar indicates the Full Width at Half Maximum of IGRINS' instrumental profile ($FWHM_{\rm{IGRINS}}$), corresponding to a minimum resolvable feature. The phase-dependent wavelength shift of the spectral lines is due to changes in winds and rotation as we see different sides of the planet's disk.} 
    \label{fig:gcm_spectra}
\end{figure}
The phase-dependent output emission spectra are complex, as they are a combination of regional absorption and emission spectral features. They are shown for four post-eclipse orbital phases in Figure~\ref{fig:gcm_spectra}. The bottom panels show the cross-sections for CO and H$_2$O which were used to identify the three spectral lines (the same as were used in the $\tau=2/3$ computation). The spectral features are Doppler broadened from winds and rotation and spectral lines are slightly Doppler-shifted from their rest frame position. A drift of the spectral lines towards longer wavelengths by $\sim$1-2 $\text{km/s}$ between the start and the end of the phase range is seen, as initially explored and mentioned by \citet{Beltz2022}.
\begin{figure*}
    \centering
\includegraphics[width=0.85\textwidth]{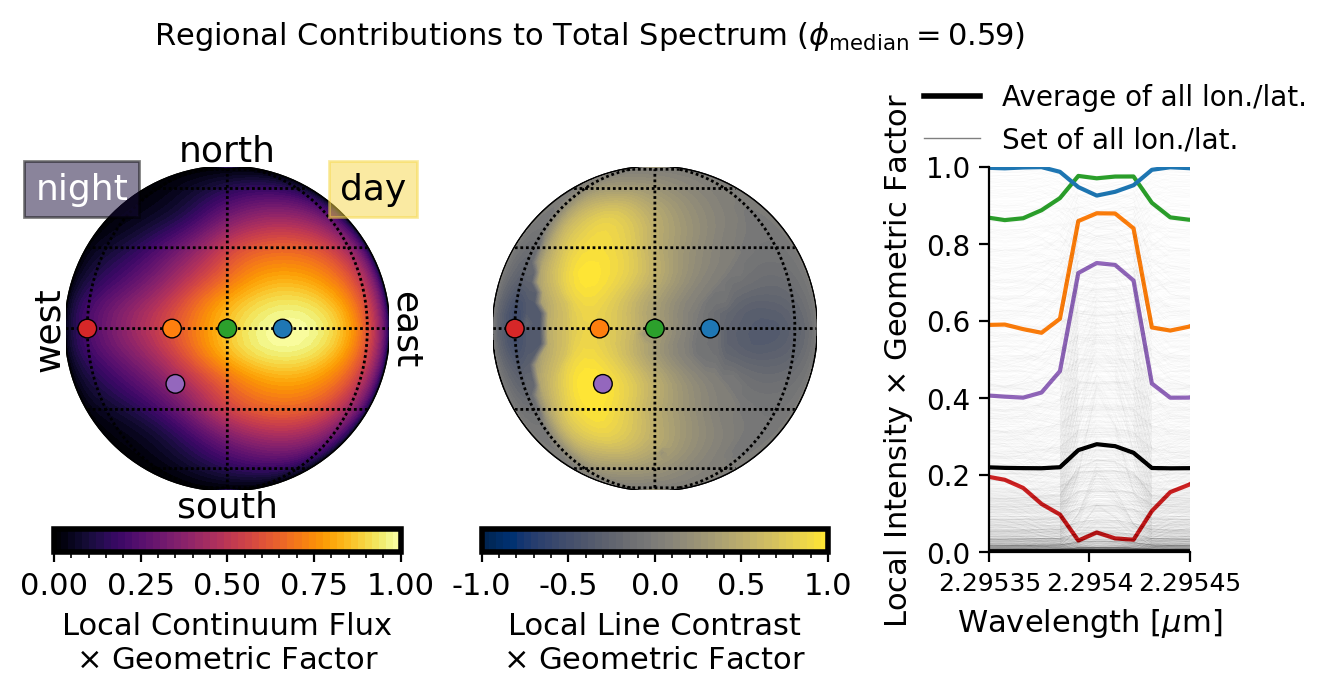} 
    \caption{Regional contributions to the total planet spectrum shown at median observed phase $\phi=0.59$, in the absence of winds and rotation. \textit{Left:} local continuum flux weighted by the geometric factor, normalized to its maximal value. The combined impact of the viewing angle and hot spot produces a peak slightly eastward. Each grid displays latitudes of 0$\degr$, $\pm30\degr$, and $\pm60\degr$, as well as the longitude at the sub-observer point, and at $\pm 60\degr$ offset from this point.  \textit{Middle:} local line contrast weighted by the geometric factor, normalized to its maximal absolute value. Some regions produce emission lines (positive values), while others produce absorption lines (negative values). \textit{Right:} Local intensity times the geometric factor. Local spectra are highlighted for five color-coded locations on the maps. They show a range of continuum flux with either emission or absorption lines. The disk-average is plotted in solid black. The set of all local spectra is plotted with the finer transparent black lines to visualize the full range.}
    \label{fig:regional_spectra} 
\end{figure*}
To further understand how each location on the planet's disk contributed to the total spectrum, we plotted the local intensity times their geometric factor Figure~\ref{fig:regional_spectra}. This is shown at a median orbital phase of $\phi_{\rm{median}}=0.59$. Thus, each local spectrum accounts for the combined effect of the thermal structure, chemical abundances, and viewing angle. Furthermore, we fitted a Gaussian profile to the CO line to obtain the continuum and line contrast at each longitude and latitude. It shows that while we receive more photons from the eastward region, this region produces shallow absorption lines. Regions further westward produce fewer photons, but stronger emission lines. The global spectrum is a weighted average of the whole disk, which has weak emission lines. This is due to the strong emission lines westward of the observer's line-of-sight being muted by the hotter regions of the atmosphere.
\section{Simulating HRS observations} \label{sec:sims}
\begin{deluxetable}{l c r}
\tablecaption{Observational setup used to create simulated high-resolution observations for IGRINS. These values are chosen such that they are typical for real HRS IGRINS post-eclipse observations. \label{table:obs_setup}}
\tablewidth{0pt}
\tablehead{
\colhead{\textbf{Parameter}} & \colhead{\textbf{Symbol}} & \colhead{\textbf{Value}}
}
\startdata
\quad Number of frames & $N_{\text{frames}}$ & 107 \\ 
\quad Number of spectral orders & $N_{\text{orders}}$ & 44 \\ 
\quad Number of pixels per order & $N_{\text{pixels}}$ & 1848 \\ 
\quad Instrumental spectral resolution & $R$ & 45,000 \\ 
\quad Relative orbital phase & $\phi$ & $0.54 - 0.64$ \\ 
\quad Continuum S/N per spectrum & $S/N$ & 200 \\
\enddata
\end{deluxetable}
To simulate HRS data, we chose a simplified observational setup that approximates real high-resolution IGRINS observations. We use a total of 107 spectra, which equals the total number of spectra collected of WASP-76 b as part of program GS-2023A-Q-222, PI: E. Rauscher. Each of the 44 orders has up to 1848 spectral channels at a spectral resolution of $R = 45,000$. We focus on observations taken during post-eclipse with an orbital phase range of $0.54 - 0.64$. We assume a continuum S/N of 200 per frame, which is typical for high-resolution emission spectroscopy, and has been achieved for other hot Jupiters with IGRINS \citep{Line2021}. If the goal was to model observational effects as realistically as possible, a constant S/N is a suboptimal approximation, especially inside the telluric line cores. However, the point here is to simply add sufficient noise to the input spectra, such that a 1D retrieval can be deployed on the dataset.
The spectral time series is built from three main components: the phase-dependent exoplanetary and phase-independent stellar and telluric spectra. The GCM output exoplanet spectra were computed at eight phases across this range, therefore we linearly interpolated between the two closest spectra in phase to obtain a spectrum for each frame. This is warranted by the relatively slow change in 3D spectra with phase (see Figure~\ref{fig:gcm_spectra}). A stellar spectrum was computed using the PHOENIX stellar library\footnote{\url{https://phoenix.astro.physik.uni-goettingen.de/}} using a stellar temperature of $T=6200 \ \rm{K}$ with no additional rotational broadening. The {\sc Telfit} telluric model was used to simulate telluric lines at a humidity of $50 \%$ and a zenith angle of $45^\circ$.
The exoplanetary spectra must be Doppler-shifted as induced by the orbital motion around its host star. To isolate the effect of velocity shifts produced by the GCM, we assume these observations have already been corrected for the barycentric velocity, implying BERV = 0 km/s. The stellar lines are assumed to be stationary, which is very close to the actual system, since the stellar radial velocity is only $\sim$0.1 km/s \citep[][]{West2016}, much smaller than the exoplanetary orbital velocity. However, detecting Doppler shifts at this level in real observations would require correction for this minimal motion. Furthermore, the orbit is circular \citep[$e < 0.05$ at 3$\sigma$,][]{West2016}, and we compute the radial velocity (RV) as
\begin{equation}
    {RV(\phi) = V_{\text{sys}} + K_{\text{p}} \sin{\phi}},
    \label{eq:rv_circular}
\end{equation}
where $V_{\text{sys}}$ is the system velocity, $K_{\text{p}}$ is the Keplerian velocity, and $\phi$ is the relative orbital phase. In our simulation, we adopted the values measured by \citet{Pelletier2023}, acknowledging that these values may reflect both the planet's orbital velocity and 3D atmospheric effects. The key is that we will compare our retrieved values against our simulated values, regardless of the specific values adopted. All spectra were also convolved to IGRINS' instrumental resolution, split into IGRINS' spectral orders, and sampled to spectral channels per order.
\begin{figure}
    \centering
    \includegraphics[width=0.49 \textwidth]{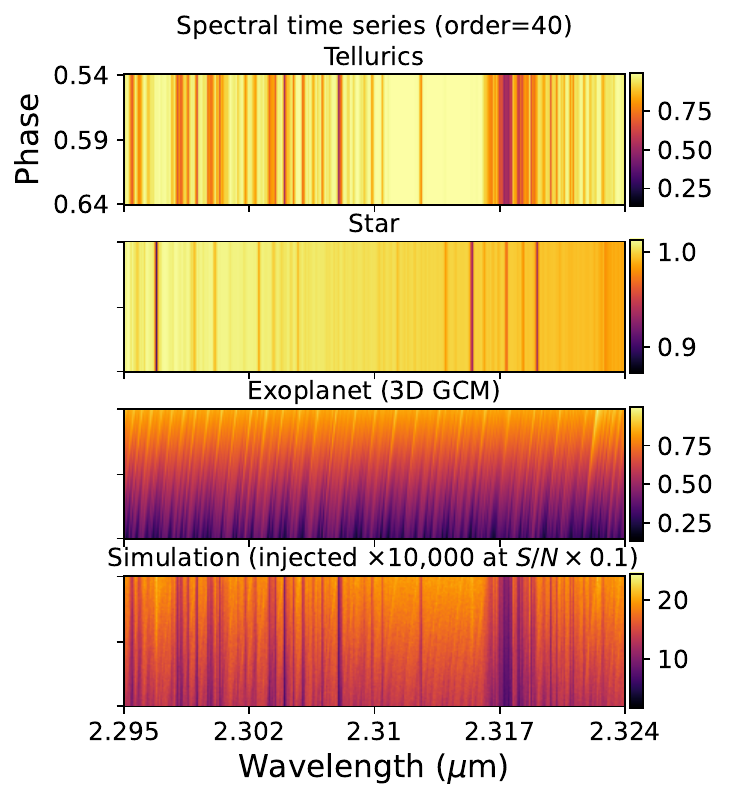} 
    \caption{The GCM spectra are used to simulate a 3D HRS data set within an observational framework. Telluric, stellar, and exoplanetary (GCM, Doppler-shifted to its orbital velocity) components used to simulate a high-resolution spectral time series which are combined using equation~(\ref{eq:inject}). Each panel shows rows of spectra as a function of the relative orbital phase. These are shown using spectral order number 40 as an example, as it covers 2.952 $\micron$ CO spectral feature. The bottom panel shows the simulated spectral time series. For visualization purposes: all panels have been normalized by their median value, the exoplanet signal has been amplified to 10,000 times its nominal strength to enhance the exoplanetary lines, and the noise has been simulated at a lower continuum signal-to-noise ratio (reduced to a tenth of the original S/N) to demonstrate the effect of white noise.}
    \label{fig:sims_steps} 
\end{figure}
The simulated high-resolution spectral time series $F(\phi)$ is computed from the Doppler-shifted exoplanetary ($F_p(\phi)$), stellar ($F_{\star}$), and telluric ($T$) spectral time series as followed:
\begin{equation}
    {F(\phi) = T \times \Big(1 + \Big(\frac{R_{\text{p}}}{R_{\star}}\Big)^2 \frac{F_{\text{p}}(\phi)}{F_{\star}} \Big) },
    \label{eq:inject}
\end{equation}
which normalizes the spectral continuum to the stellar continuum. A Poisson noise matrix is added to the spectral time series, at the continuum $S/N = 200$ level. An overview of each component and the final simulated spectral time series is shown in Figure~\ref{fig:sims_steps}.
The spectral time series is reduced using Principal Component Analysis (PCA). This mimics the data post-processing used for real IGRINS observations to remove the quasi-stationary telluric and stellar components. We remove two principal components (PCs), a relatively small number, to avoid degrading the exoplanetary signal, while still doing all the same post-processing steps as on real HRS observations. This choice is warranted by a recent investigation of the error budget in HRS which suggests that removing a small number (PCs$\leq4$) of PCs does not significantly degrade the exoplanetary signal such that retrieved atmospheric conditions are impacted \citep{Savel2024}. We fixed the number of PCs to use, instead of determining the 'optimal number' of PCs to remove, as this is non-trivial and no one robust metric or method exists to date to determine this \citep{Cabot2020, Cheverall2023, Spring2022, Smith2024}.
\begin{deluxetable*}{p{3cm} p{2cm} p{2.5cm} p{2.5cm} p{2.5cm}}
\tablecaption{Summary of the 1D retrievals on the simulated 3D HRS dataset. The retrieved constraints with/without water dissociation and with/without rotational broadening are broadly consistent with each other. \textit{First column:} lists the parameters used in the analyses. The CO and H$_2$O VMR are $X_{\rm{CO}}$ and, respectively, $X_{\rm{H_2O}}$. The P-T parameters are from the \citet{Madhusudhan2009} parameterization: $P_1$, $P_2$, and $P_3$ set the boundary pressures of three atmospheric layers, $\alpha_1$ and $\alpha_2$ set the thermal gradient in the first and second layers, and $T|_{P = 1\, \text{\rm{$\mu$bar}}}$ sets the temperature at the top of the atmosphere. The offsets from the Keplerian and system velocity are $\Delta K_{\rm{p}}$ and $\Delta V_{\rm{sys}}$. The FWHM of the rotational broadening kernel is $\delta V_{\rm{broad}}$. \textit{Second column:} shows the prior range explored. \textit{Third column:} The retrieved values for each experiment are shown in the three rightmost columns. The experiments either included water dissociation ($\rm{H_2O}$ diss.) or a constant vertical water abundance was used (No $\rm{H_2O}$ diss.). Rotational broadening was either included as a free parameter (Rot. broad.) or excluded (No Rot. broad.). \label{table:retrieval}}
\tablewidth{0pt}
\tablecolumns{5}
\tablehead{
\colhead{\textbf{Parameter}} & \colhead{\textbf{Prior range}} & \multicolumn{3}{c}{\textbf{Retrieved values}} \\
\colhead{} & \colhead{} & \colhead{H$_2$O diss.} & \colhead{H$_2$O diss.} & \colhead{No H$_2$O diss.} \\
\colhead{} & \colhead{} & \colhead{Rot. broad.} & \colhead{No Rot. broad.} & \colhead{Rot. broad.}
}
\startdata
\textbf{Chemistry} \\ \hline
\quad $\log(X_{\mathrm{CO}})$ & $-9 \rightarrow -1$ & $\rm{-4.0^{+0.8}_{-0.6}}$ & $\rm{-4.3^{+0.7}_{-0.6}}$ & $\rm{-4.0^{+0.8}_{-0.6}}$ \\
\quad $\log(X_{\mathrm{H_2O}})$\tablenotemark{1} & $-9 \rightarrow -1$ & $\rm{-4.6^{+0.6}_{-0.6}}$ & $\rm{-4.9^{+0.6}_{-0.6}}$ & $\rm{-4.6^{+0.7}_{-0.6}}$ \\ \hline
\textbf{P-T profile} \\
\quad $T|_{P = 1\, \text{\rm{$\mu$bar}}}$ (K) & $1500 \rightarrow 4000$ & $\rm{2232^{+603}_{-434}}$ & $\rm{2045^{+629}_{-376}}$ & $\rm{2171^{+558}_{-378}}$ \\
\quad $\log(P_1 \, (\mathrm{bar}))$ & $-5 \rightarrow 2$ & $\rm{-3.9^{+0.9}_{-0.7}}$ & $\rm{-3.4^{+1.5}_{-1.1}}$ & $\rm{-3.8^{+1.0}_{-0.8}}$ \\
\quad $\log(P_2 \, (\mathrm{bar}))$ & $-5 \rightarrow 2$ & $\rm{-0.8^{+0.4}_{-0.5}}$ & $\rm{-0.5^{+0.4}_{-0.4}}$ & $\rm{-0.8^{+0.4}_{-0.5}}$ \\
\quad $\log(P_3 \, (\mathrm{bar}))$ & $-2 \rightarrow 2$ & $\rm{1.2^{+0.5}_{-0.6}}$ & $\rm{1.3^{+0.4}_{-0.5}}$ & $\rm{1.3^{+0.5}_{-0.6}}$ \\
\quad $\alpha_1 \, (\mathrm{K^{-1}})$ & $0.05 \rightarrow 0.95$ & $\rm{0.59^{+0.23}_{-0.24}}$ & $\rm{0.66^{+0.20}_{-0.26}}$ & $\rm{0.62^{+0.23}_{-0.25}}$ \\
\quad $\alpha_2 \, (\mathrm{K^{-1}})$ & $0.05 \rightarrow 0.95$ & $\rm{0.25^{+0.03}_{-0.03}}$ & $\rm{0.28^{+0.05}_{-0.05}}$ & $\rm{0.26^{+0.03}_{-0.03}}$ \\ \hline
\textbf{Velocities} \\ \hline
\quad $\Delta K_{\rm{p}} \, (\mathrm{km \, s^{-1}})$ & $-25 \rightarrow 25$ & $\rm{1.2^{+4.4}_{-4.6}}$ & $\rm{-0.9^{+7.5}_{-5.0}}$ & $\rm{1.2^{+4.5}_{-4.7}}$ \\
\quad $\Delta V_{\mathrm{sys}} \, (\mathrm{km \, s^{-1}})$ & $-25 \rightarrow 25$ & $\rm{-1.4^{+2.6}_{-2.6}}$ & $\rm{-3.1^{+5.3}_{-3.0}}$ & $\rm{-1.4^{+2.5}_{-2.7}}$ \\
\enddata
\tablenotetext{1}{Or $\log(X_{\mathrm{0,H_2O}})$, the water abundance at the deepest pressure layer ($P = 10^{2.5} \, \text{bar}$), for the experiments with dissociated H$_2$O.}
\end{deluxetable*}
\vfill\eject
\section{1D Retrieval: Setup and Results} \label{sec:retrieval}
\subsection{1D retrieval setup}
A retrieval framework systematically compares data to a set of models using a Bayesian estimator. The data here is the simulated spectral time series from Section~\ref{sec:sims}. Each model is a spectral time series of an exoplanetary spectrum divided by the stellar spectrum, where the exoplanetary spectrum is computed by running a 1D forward model, hence our retrieval framework is 1D.
The 1D forward model computes an exoplanet emission spectrum given chemical abundances, and temperature, as a function of pressure. The forward model used in this work is adapted from the 1D retrieval framework introduced by \citet{Line2021}. This forward model creates 107 logarithmically equally-spaced atmospheric layers between pressures of $10^{-6}-10^{2.5} \ \text{bar}$. In each atmospheric layer, chemical abundances and temperatures are defined. It assumes a black body flux is emitted from the deepest layer radiating upwards. The output of each subsequent layer is calculated using the radiative transfer equations \citep{Line2013, Line2017, Brogi2019}, thus accounting for absorption and emission processes due to each opacity source in each layer. The observed emitted spectrum is the flux radiated outwards in the topmost layer of the modeled atmosphere.
In our forward model setup, we use parameterized P-T and chemical profiles. We adopt the P-T parameterization by \citet{Madhusudhan2009}, due to its flexibility to fit a wide range of both inverted and non-inverted profiles. In principle, alternative P-T parameterizations could be considered. We will revisit the choice of P-T profile in Section~\ref{sec:results} and~Section~\ref{sec:discussion_caveats}. For the chemical profiles, our default setup assumes constant vertical abundances. However, the Volume Mixing Ratio (VMR) can deviate several orders of magnitude from constant vertical abundances at lower pressures depending on the P-T structure. Therefore, following \citet{Gandhi2024}, we also run forward models with a parameterized water chemical profile
\begin{equation}
    {\frac{1}{\sqrt{X_{\rm{H_2O}}(P, T)}} = \frac{1}{\sqrt{X_{\rm{0,H_2O}}}} + \frac{1}{\sqrt{X_{\rm{d,H_2O}}(P, T)}}}.
    \label{eq:water_dissociation}
\end{equation}
Here $X_{\rm{0,H_2O}}$ is the water abundance at the deepest pressure layer and $X_{\rm{d,H_2O}}$ is given by
\begin{equation}
    {\log X_{\rm{d,H_2O}}(P, T) = 2 \log(P)+\frac{4.83 \times 10^4}{T} - 15.9,}
\end{equation}
which follows the parameterization by \citet{Parmentier2018}.
The atmospheric composition includes hydrogen, helium, CO, and H$_2$O. The opacity line lists for H$_2$O and CO are the same as those used in the GCM. We assumed a mixture of hydrogen and helium with a number ratio $n_{\rm{H_2}}/n_{\rm{He}} = 0.176471$ following \citet{Line2021}. The mean molecular weight is calculated by adding the contributions of hydrogen, helium, CO, and H$_2$O accordingly. Finally, the radiative transfer equations require the gravitational acceleration $g$, which in the case of WASP-76 b is $\log g = 2.83$ (in cm$^2$/s) (see Table~\ref{table:w76_system_params}). 
The forward modeled 1D emission spectra are computed at $R = 125,000$. For each frame, the computed spectrum is Doppler-shifted by a system velocity $V_{\text{sys}} + \Delta V_{\text{sys}}$ and a Keplerian velocity $K_{\text{p}} + \Delta K_{\text{p}}$, following equation~(\ref{eq:rv_circular}) and where
where $K_{\text{p}}$ and $V_{\text{sys}}$ are the values reported in Table~\ref{table:w76_system_params}.
These spectra do not yet account for instrumental or rotational broadening. The spectra are convolved to a lower resolution using a Gaussian kernel to account for IGRINS' instrumental resolution of $R = 45,000$. Convolution with an additional rotational broadening kernel $\delta V_{\rm{broad}} \equiv V_{rot} \sin{i}$ can be applied. This is a computationally inexpensive way to approximate in parts the effect of rotational broadening on the spectrum, despite the GCM spectra having more complex line shapes. We follow the approach by \citet{Finnerty2024} and use the rotational kernel by \citet{Carvalho2023}.
Previous work by \citet{Line2021} and \citet{Gibson2022} cautioned against the direct comparison of the forward model spectrum to the residual (post-PCA) spectral time series. This is because, despite PCA predominantly removing the telluric lines, stellar lines, and systematic trends, it also erodes part of the exoplanetary signal in the process. To account for this, the discarded eigenvectors from the SVD of the observed spectral time series are used to reconstruct a matrix capturing all of the principal trends. An injected spectral time series is computed by matrix multiplication of the reconstructed matrix with $(1+F_{\text{p}}(\phi)/F_{\text{s}})$. PCA is re-applied to the injected spectral time series. Finally, the reprocessed spectra are compared to the post-PCA data spectra.
The Bayesian estimator used in this work is the log-likelihood from \citet{Brogi2019}, defined as
\begin{equation}
    {\log{L} = -\frac{N}{2}\log{\bigl( s_{F}^2 - 2 R + s_{F_{\text{m}}}^2 \bigl)}}
    \label{eq:logL_brogi2019}
\end{equation}
where $F$ is a post-PCA data spectrum, $F_{\text{m}}$ a reprocessed spectrum, $R = cov(F, F_{\text{m}})$ the cross-covariance between them, and $N$ the total number of spectral channels. Sometimes an additional nuisance scale parameter, a scalar $a$ \citep[first introduced by][]{Brogi2019} is included within the retrieval to scale the model spectrum. In our framework, we explicitly omit it (or equivalently set $a=1$), as it can scale the line contrast to mimic changes in the thermal gradient or chemical abundances in non-trivial ways, leading to additional complexity in understanding the retrieved result and possibly a bias to the retrieved solution. We get back to this point in Section~\ref{sec:discussion_caveats}.
The {\sc Python} Bayesian nested sampling algorithm package {\sc PyMultinest}(version 2.12) was used to iteratively compute forward models, reprocess the models, and compute the log-likelihood. We used a total of 400 live points. The computational time of each forward model was reduced by using a single NVIDIA Tesla V100 GPU to perform the radiative transfer calculations. This is because GPUs can be faster than conventional CPUs for computations that can run in parallel, such as our radiative transfer algorithm. The model reprocessing time was also reduced by parallelization for each spectral order on 24 CPUs. We found most 1D retrievals took about $\sim$1-2 days to complete using this setup, depending on the number of free parameters included.
To verify the retrieval algorithm was implemented correctly, we first ran a 1D mock retrieval on a 1D HRS dataset, which is a dataset that was simulated using the retrieval's 1D forward model. In this case, the retrieval algorithm correctly retrieved the P-T and chemical profiles within 1$\sigma$ to 2$\sigma$. This is described in more detail in Section~\ref{sec:mock_retrieval} of the appendix.
We run 1D retrieval experiments on the 3D GCM phase-dependent HRS datasets. To explore the effects of the 1D retrieval's parameterization of water dissociation and rotational broadening we run three distinct retrieval experiments: (1) with dissociated water and including rotational broadening (2) with dissociated water and excluding rotational broadening, and (3) no dissociated water (constant vertical abundance) and including rotational broadening.
\subsection{1D retrieval results on 3D GCM simulated data} \label{sec:results}
\begin{figure}
    \centering
    \includegraphics[width=0.45 \textwidth]{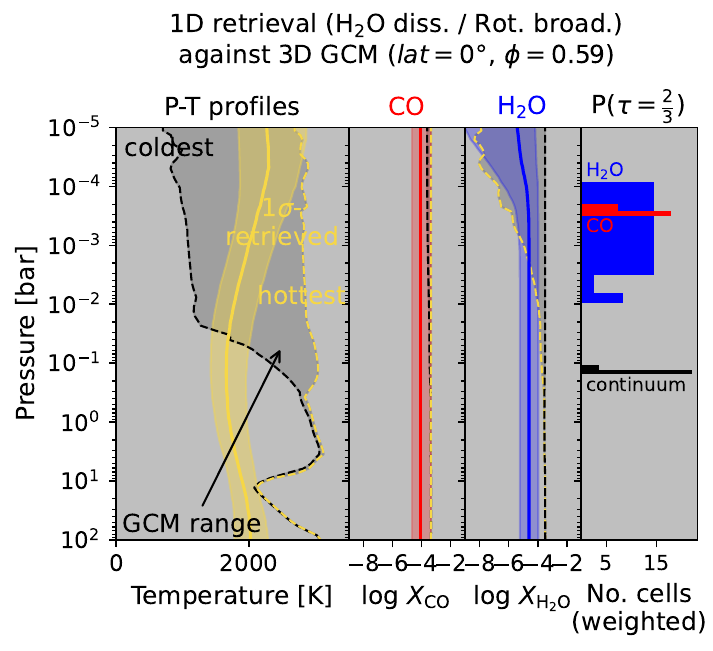}
    \includegraphics[width=0.45 \textwidth]{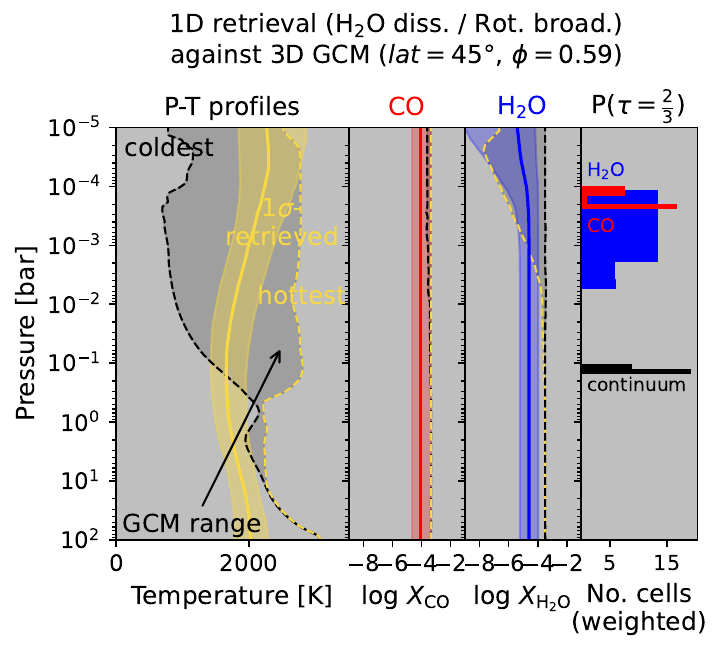}
    \caption{Retrieved 1D P-T and chemical profiles on the 3D GCM HRS dataset show atmospheric conditions are broadly retrieved, although chemical abundances are slightly underestimated. Results are shown for the retrieval parameterization with dissociated $\rm{H_2O}$ and including rotational broadening. The panels show the range of P-T and chemical profiles in the GCM that is visible to the observer at mean orbital phase $\phi_{\rm{mean}}=0.59$. The top panel shows the GCM along latitudes of $0\degr$ and the bottom panel along latitudes of $45\degr$. The coolest and hottest profiles are indicated by the dashed black and yellow lines, respectively. The horizontal dark blue, dark red, and dark gray regions indicate the range of pressures probed by water lines ($\tau_{\rm{H_2O}}=\frac{2}{3}$), CO lines ($\tau_{\rm{CO}}=\frac{2}{3}$), and the spectral continuum ($\tau_{\rm{cont}}=\frac{2}{3}$) respectively. The 1D retrieved profiles are shown with solid lines and their 1$\sigma$-confidence intervals are shaded; the retrieved P-T profile is shown in yellow, the retrieved CO abundance profile is the vertical shaded region in red, and the retrieved $\rm{H_2O}$ abundance profile is the vertical shaded region in blue.}
    \label{fig:retrieval_to_gcm_3D}
\end{figure}

The 1D retrieved parameters on the 3D simulated HRS dataset are summarized in Table~\ref{table:retrieval}. The corresponding marginal distributions (colloquially referred to as corner plot) are included for completeness in Section~\ref{sec:corner_plots} of the appendix. The key thing to take away is that they are all consistent within 1$\sigma$ with each other. This suggests rotational broadening and water dissociation have a negligible effect on constraints of the chemical bulk abundances and P-T structure, at least for this specific simulation setup of WASP-76 b.
The 1D retrieval constraints for the parameterization with dissociated water and rotational broadening are compared to the GCM in Figure~\ref{fig:retrieval_to_gcm_3D}. Two panels are used to visualize the range of thermal and chemical profiles in the GCM, for longitudinal profiles visible at $\phi_{\rm{median}}=0.59$: one panel shows equatorial latitude and the other panel shows $+45\degr$ latitude. The leftmost panel shows that the retrieved P-T profile is broadly within the range of conditions present in the atmosphere. The profile is significantly cooler than the hottest dayside profiles, particularly at deeper pressure levels for reasons we discuss below. Both the retrieved CO and H$_2$O abundances are slightly lower compared to values expected from the GCM on both the dayside and nightside.

\section{Discussion} \label{sec:discussion}
\subsection{Sensitivity to longitudinal P-T profiles} \label{sec:discussion_p-t}
Understanding how the retrieved 1D P-T structure compares to the 3D P-T structures in the GCM is non-trivial. This is because our retrieval compares a forward-modeled 1D spectrum to inherently phase-dependent 3D spectra. Moreover, the HRS methodology removes the absolute continuum of the spectra in our data processing.
\begin{figure*}
    \centering
    \includegraphics[width=0.47\textwidth]{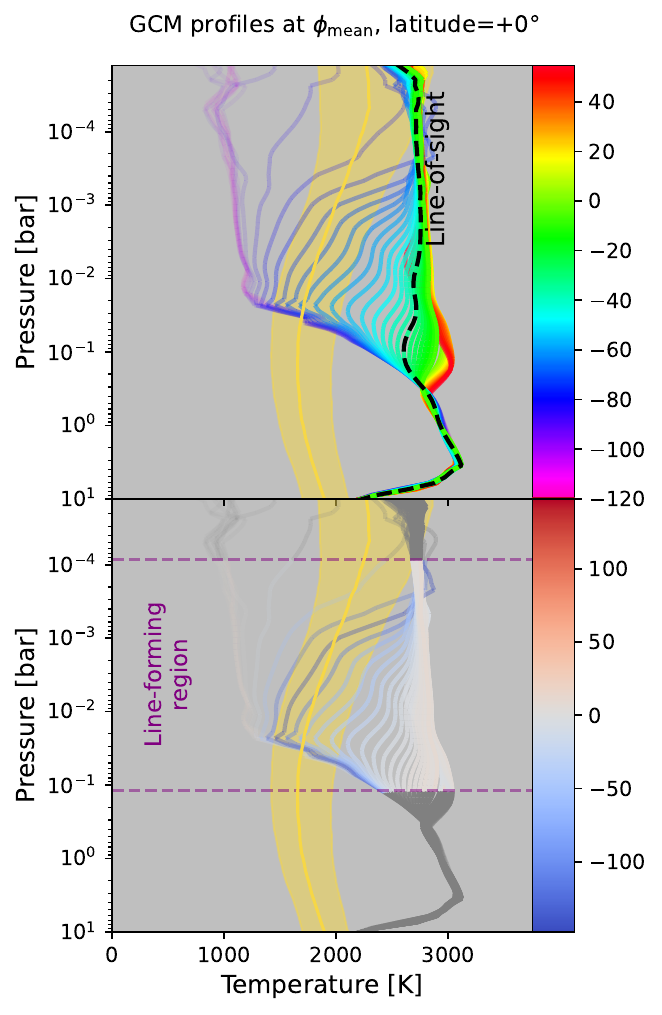} 
    \includegraphics[width=0.49\textwidth]{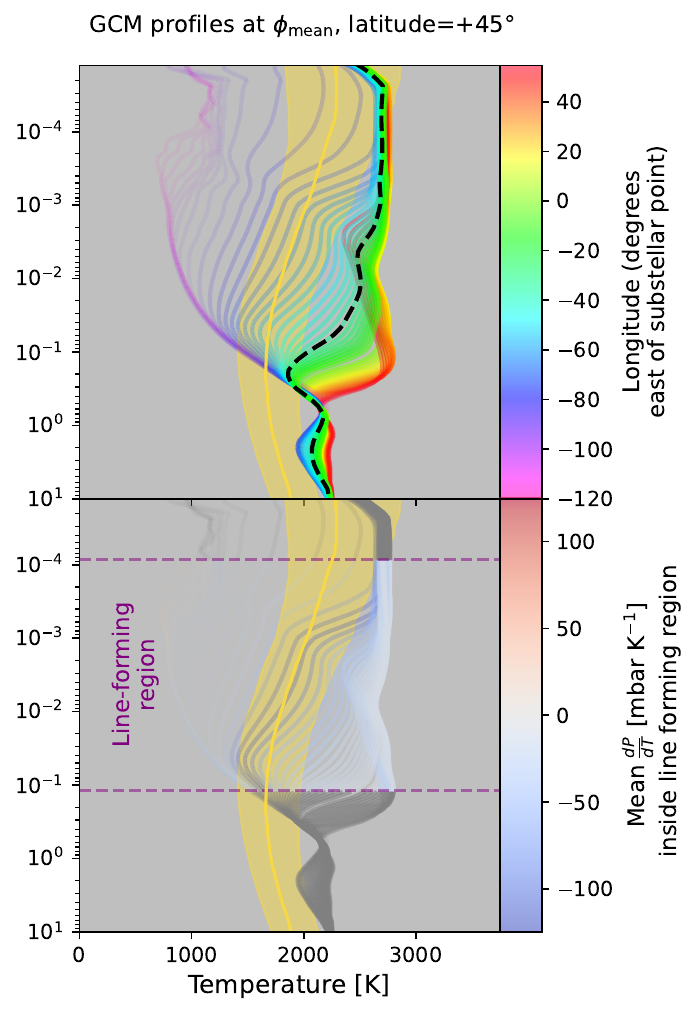}
    \caption{Comparison of the GCM P-T profiles and the 1D retrieved P-T profile to show we are more sensitive to the westward sides of the exoplanet's disk, which have the strongest inversions in the spectral line-forming region. The retrieved 1D profile is the solid yellow line, where the shaded region indicates the region within the 1$\sigma$ confidence interval. This is for the retrieval experiment that includes water dissociation and rotational broadening. \textit{Top panels:} shows the P-T profiles along latitudes of $0\degr$ and $45\degr$ visible at the mean orbital phase $ \phi_{\text{mean}}=0.59$. The color indicates the longitude in degrees eastward from the sub-stellar point. At this phase, the line-of-sight is centered at a longitude of $-32^\circ$ (dashed black line), thus more of the westward side of the planet is visible. \textit{Bottom panels:} the same GCM P-T profiles, but here the colors indicate the strength of the mean local temperature gradient ($\frac{dP}{d{T}}$) over the line-forming region. We define the line-forming region (shaded magenta) as the region between the lowest pressure where either CO or H$_2$O spectral line cores form and the highest pressure where the spectral continuum forms. The 1D retrieved profile aligns well with those profiles that have the most negative mean temperature gradient, corresponding to the GCM P-T profiles with the strongest net inversions inside the line-forming region.}
    \label{fig:gcm_longitudes}
\end{figure*}
To interpret what regions of the exoplanetary disk we are most sensitive to, we compare the retrieved P-T profile to a set of longitudinal P-T profiles in the GCM. This is shown in the top panels of Figure~\ref{fig:gcm_longitudes}. Plotted are a set of longitudes at latitudes $0\degr$ and $45\degr$ when observed at the mean orbital phase $\phi=0.59$. P-T profiles along longitudes in the hot spot region (eastward coinciding with positive longitudes) are more isothermal or even slightly non-inverted, in line with the regional spectra shown in Figure~\ref{fig:regional_spectra}. The hot spot region generally has smaller temperature gradients than regions towards the edge of the disk (westwards, coinciding with negative longitudes).
The retrieved 1D P-T profile is closest in temperature to the GCM profiles corresponding to the cooler westward regions around the edges of the disk, which visually corresponds to longitudes $\sim 50^\circ-60^\circ$ westward from the sub-stellar point. This is not just a geometrical effect of seeing more of the west side of the planetary disk at $\phi=0.59$ as our line of sight is centered at a longitude of $32^\circ$ westward from the sub-stellar point.
To understand this, we also plot the mean temperature gradient inside the line-forming region for each GCM P-T profile in the bottom panels of Figure~\ref{fig:retrieval_to_gcm_3D}. This value was computed by taking the set of temperature gradients at each P-T point inside the line-forming region and computing the mean of this set. We define the line-forming region as the region between the pressure where the CO or H$_2$O line cores form and the pressure of the spectral continuum, which is set by the range of P($\tau=2/3$) values in the histograms presented earlier in Figure~\ref{fig:gcm_longitudes}. The retrieved 1D P-T profile aligns well with those GCM P-T profiles that have the largest absolute temperature gradients within the line-forming region. They correspond to those parts of the atmosphere that have a net inversion throughout most of the line-forming region. A comparison between latitudes of $0\degr$ and $+45\degr$ may explain why the 1D P-T profile does not capture the change from inverted to non-inverted profiles $\sim50-100$ mbar. The profiles stay non-inverted at those pressures for latitudes of $45\degr$. Even though these regions are seen at an increased angle with the line-of-sight, they maximize the local line contrast as was shown in Figure~\ref{fig:regional_spectra}. An alternative explanation may be that the lack of flexibility of the P-T parameterization, which we will get back to.
\begin{figure*}
    \centering
    \includegraphics[width= 0.9\textwidth]{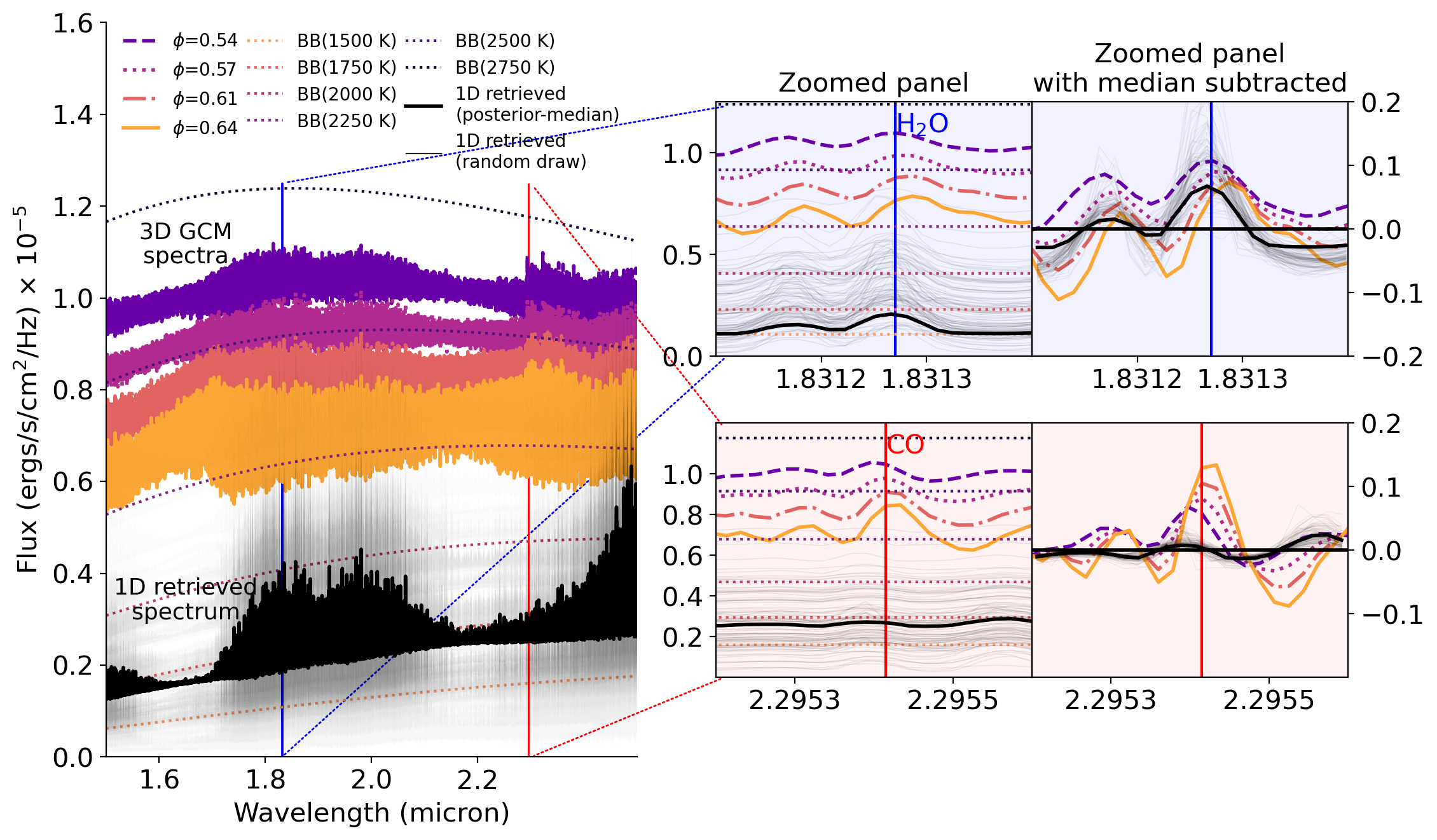}
    `\caption{3D GCM phase-dependent spectra compared to the 1D retrieved posterior-mean spectrum demonstrate we are offset in absolute flux, but broadly consistent in terms of spectral line contrast of the water lines. Results are shown for the retrieval that included rotational broadening and water dissociation. To understand the range of retrieved 1D spectra given our posteriors, we over-plot 1D spectra for a hundred random draws using a reduced linewidth and transparency. \textit{Left panel:} Four GCM phase-dependent spectra are plotted in color, and the 1D retrieved posterior-median spectrum is shown in solid black. Additionally, black bodies are plotted at various temperatures with a dotted line style. \textit{Middle panel:} a zoomed panel on the same wavelength range as shown in Figure~\ref{fig:gcm_spectra}. A H$_2$O line and CO line are highlighted using a vertical blue and red line. \textit{Rightmost panel:} the zoomed panel with the median of each spectrum subtracted.}
 \label{fig:dopon_spectra_comparison}
\end{figure*}
To understand why large temperature gradients seem favored, we compared the retrieved posterior-median spectrum to the GCM 3D phase-dependent spectra. A comparison of the GCM phase-dependent spectra to the retrieved spectrum is shown in Figure~\ref{fig:dopon_spectra_comparison}. Because HRS removes the continuum, we also compare the spectra after subtraction of the median in the right-most panel. To estimate the range set by the retrieved posteriors, we also over-plot a set of a hundred randomly drawn 1D spectra.
Comparison of the retrieved 1D spectrum to the GCM spectra demonstrates we are sensitive to the relative line contrast, particularly of the water lines, but not to the spectral continuum. This is highlighted by the random draws, which show the continuum is broadly unconstrained. This is in line with results from previous HRS studies \citep[e.g.][]{Brogi2019}. For this GCM, we find the 1D retrieved spectrum is significantly offset in terms of absolute flux from the phase-dependent spectra. The 1D retrieved spectral continuum has brightness temperatures about eight times cooler than the GCM spectra. However, after post-processing the high-resolution spectra, the continuum is removed. In the rightmost panel, we can see that the 1D retrieved spectrum broadly matches the relative spectral H$_2$O lines. However, especially the CO lines are too shallow. At the same time, the retrieved chemical abundance of CO is too low compared to the 3D GCM. Although the continuum level is largely unconstrained, lower continuum fluxes are favored compared to those in the GCM spectra. We speculate this may result from the line cores determining the maximum temperature, combined with the limited flexibility of the \citet{Madhusudhan2009} P–T parameterization to capture steeper thermal gradients at higher pressures.
Degeneracies between the P-T profile and the chemical abundances can be seen in Figure~\ref{fig:corner_3D_dopoff_H2O_dissociated} as the correlation between $\log P_2$ and the chemical abundances. As shown in Figure~\ref{fig:gcm_longitudes}, the water and CO line cores form at different pressure levels, thus sampling various parts of the P-T profile. Inversions are generally steeper around CO line forming pressures than H$_2$O line core forming pressures. However, the \citet{Madhusudhan2009} P-T parameterization only allows for a single parameter describing the steepness of the inversion layer. Moreover, this also sets the steepness of the thermal gradient of the non-inversion in the deeper atmosphere, possibly explaining why we are not matching the deeper parts of the P-T profile just above the continuum pressures. Perhaps, more flexible 1D P-T profiles, such as descriptions using a set of free-floating P-T points that are connected \citep[e.g.][]{Bazinet2024}, could improve upon this.
Since we are matching the relative spectral features in the global spectrum, we are sensitive to the complex interplay between the local continuum flux, line contrast, and disk geometry. This is because a larger temperature gradient will set a larger temperature difference between where the line cores form and where the spectral continuum forms. We emphasize these are not necessarily the hottest regions in the atmosphere, which in this GCM of WASP-76 b, corresponds with the hot spot having a weak temperature gradient (isothermal/slightly non-inverted). This is in line with the argument made to explain observational evidence of phase-dependent variation of CO emission lines seen in the atmosphere of WASP-33 b \citep{vanSluijs2023}. At the same time, hotter regions produce more photons overall, and fewer photons reach us from regions towards the edge of the disk, which are seen at a larger angle of incidence. The net result in this GCM is a global spectrum with weak emission lines (shown in Figure~\ref{fig:regional_spectra}). This may explain why the retrieved thermal profile aligns with the westward regions, which produce stronger emission lines, in combination with lower chemical abundances and shallower thermal gradient, which mute the spectral lines.
The fact that HRS retrievals seem sensitive to the disk-averaged line contrast but not the overall continuum may have important implications when jointly fitting space- and ground-based observations. Previous studies \citep{Brogi2018, Brogi2019, Smith2024} have shown that combining high-resolution and low-resolution data can improve constraints on the 1D P-T profile because they provide complementary information; the high-resolution data provides the detailed line shape, whereas the low-resolution data preserves the broader continuum information. However, these studies used a single 1D P-T profile to fit both the high-resolution and low-resolution data. This assumption may be reasonable for those exoplanets where the region of largest spectral feature formation broadly coincides with the region of greatest absolute flux. We speculate that when those regions are distinct, as in the 3D GCM explored in this work, the high-resolution and low-resolution data may have unequal sensitivity to different parts of the exoplanet. In that case, a joint fit with a single P-T profile will likely not be able to capture this fully.
Similarly to how the P-T structure is inhomogeneously sampling spatial and phase-dependent atmospheric conditions, the same would be true for the chemical profiles. In that case, one may expect to retrieve chemical abundances closer to the chemical profiles of the westward regions. As shown in Figure~\ref{fig:gcm_maps}, the westward regions have a slightly higher chemical abundance of CO ($\Delta log_{\rm{CO}} \sim +0.2$) and a higher H$_2$O abundance ($\Delta log_{\rm{H_{2}O}} \sim +1$). Contrary to this hypothesis, our retrieved chemical abundances are too low. Instead, we suggest that previously reported degeneracies between the P-T profile parameters and chemical abundances are a more plausible explanation for why the retrieved chemical abundances are too low.
\subsection{Water dissociation}
\begin{figure}
    \centering
    \includegraphics[width=0.49 \textwidth]{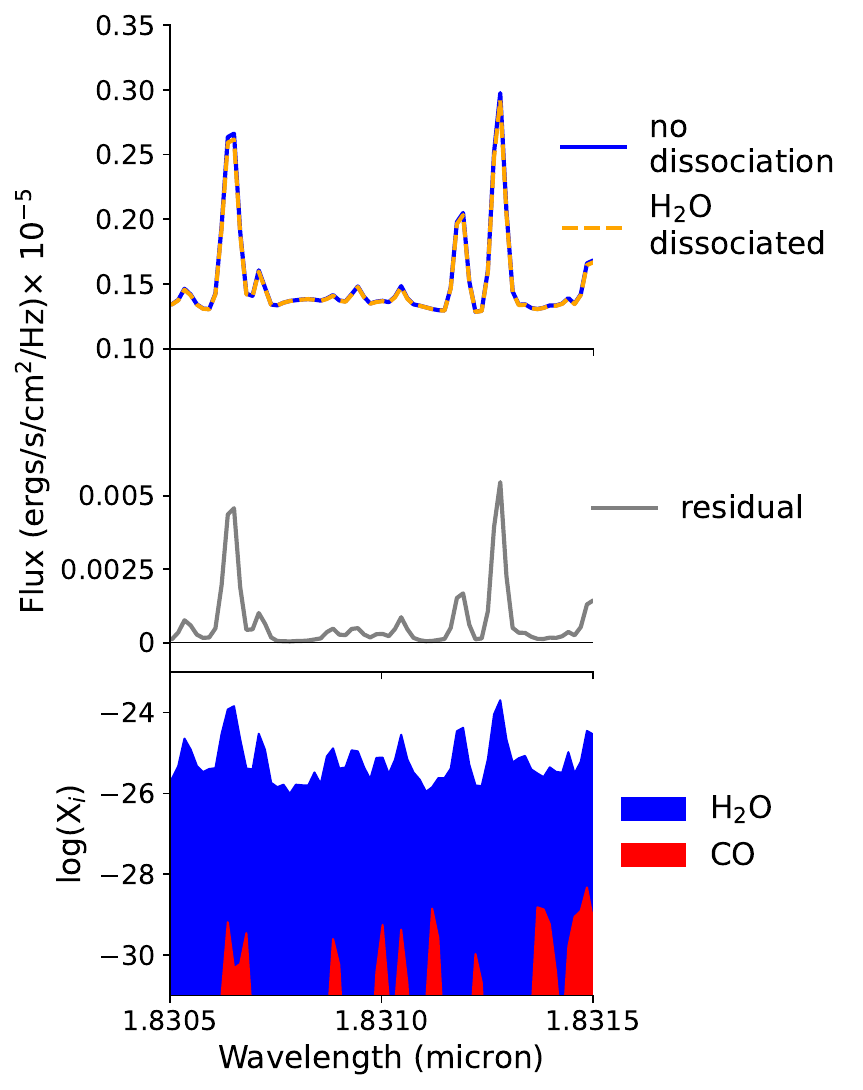}
    \caption{1D spectra with and without water dissociation to show there is a negligible difference between them for our 1D retrieved P-T range. \textit{Top:} 1D spectrum generated using the same P-T profile, H$_2$O, and CO bulk atmospheric abundance, but with (dashed orange line) or without water dissociation (solid blue line) included. \textit{Middle:} residual (solid gray line) when subtracting the model with H$_2$O dissociated from the model without it. \textit{Bottom:} range of CO and H$_2$O cross-section for temperatures between 1500-2500 K at 0.1 mbar. This demonstrates these are H$_2$O lines and not CO. For WASP-76 b, the spectra are almost within $10^{-7}$ in flux, and the effect of water dissociation is negligible.}
    \label{fig:water_dissociation}
\end{figure}
In our 1D retrieval, including or excluding H$_2$O dissociation does not significantly impact retrieved chemical abundances or P-T profiles. This is seen by comparing the values in Table~\ref{table:retrieval} between the retrieval results with and without H$_2$O dissociated, which are within 1$\sigma$ of each other. Consequently, as shown in Figure~\ref{fig:water_dissociation}, the spectra with and without dissociated H$_2$O are almost identical. Additionally, as discussed in section~\ref{sec:discussion_p-t}, we are sensitive to the cooler west-side dayside regions on the exoplanetary disk. As shown in the rightmost panel of Figure~\ref{fig:retrieval_to_gcm_3D}, these regions are closer to the non-dissociated nightside chemical profiles, which have almost constant vertical chemical abundance. Thus, the parameterization choice of water dissociation does not significantly impact retrieval results. This insensitivity to water dissociation for our simulated emission spectroscopy differs from previous findings from transmission spectroscopy by \citet{Gandhi2024}, who found water dissociation did impact retrieved results from observations of WASP-76 b. The difference is likely attributed to the different pressures probed, since transmission spectroscopy is generally more sensitive to the upper parts of the atmosphere where water dissociation is significant.
\subsection{Rotation and winds}
\label{sec:rotation_winds}
Spectral features are broadened and shifted by rotation and winds. In emission spectra, the amount of broadening and the presence of any net Doppler shift will depend on how much regions with different line-of-sight velocities contribute to the disk-integrated spectrum \citep{Zhang2017}. For example, suppose the disk-integrated spectrum is dominated by a signal from a planet region where rotation and winds bring material toward the observer. In that case, the spectral features will have a net blue shift. These effects of rotation and winds on spectral features have been explored previously in the literature. Velocity offsets and broadening have commonly been observed using transmission spectroscopy \citep[e.g.][]{Ehrenreich2020, Zhang2022, Casasayas-Barris2022, Simonnin2024} and emission spectroscopy \citep[e.g.][]{Brogi2023, Finnerty2024, Lesjak2024}. GCMs also predict species-dependent velocity offsets in transmission and emission spectroscopy \citep{Beltz2024, Wardenier2025}, although these models struggle to match the magnitude of observed velocity offsets quantitatively \citep[e.g.][]{Brogi2023}.
We retrieved a rotational broadening of $\delta v_{\rm{rot}} \sin{i} = 6.5^{+1.3}_{-1.5} \ \text{km/s}$. If WASP-76 b rotated like a solid body, assuming a tidal synchronization, the expected broadening would equal $v_{\rm{rotation}} = 5.3 \ \rm{km/s}$. A previous study using a slightly different version of our WASP-76 b GCM, e.g. using 'double gray' mode rather than the 'picket fence' mode used in this work, finds expected broadening velocities of $\pm 8-10 \ \rm{km/s}$ from the combined solid body rotation, jets and day-to-night side winds \citep[][see Figure 2]{Beltz2022}. In summary, our value is consistent (within 1$\sigma$) with a solid body rotation but underestimates the effects of jets and day-to-nightside winds. Since especially H$_2$O is not homogeneously distributed, we hypothesize its spectral features originate mostly on one side of the disk, resulting in a narrower broadening profile. This is because they are the weighted average of contributions from the full exoplanet disk. We have already seen that the features do not originate from the regions with near-isothermal P-T profiles, and overall, less flux is coming from the cooler nightside.
Based on the rightmost panels of Figure~\ref{fig:dopon_spectra_comparison}, we suggest that small wavelength offsets between the center of the H$_2$O and CO lines may have caused the retrieved chemical abundances to be too low for CO. As reported, we found a $\Delta v_{\rm{sys}} = -1.4^{+2.6}_{-2.6} \ \rm{km/s}$ and $\Delta K_{\rm{p}} = 1.2^{+4.4}_{-4.6} \ \rm{km/s}$, which is within 1$\sigma$ of the exoplanet's system and Keplerian velocities. In the right-most panels of Figure~\ref{fig:dopon_spectra_comparison}, one can see that the small velocity offsets align the retrieved spectrum with the centroid of the water lines. However, the same velocity offset results in a misalignment of the CO lines in the 1D spectrum compared to the 3D spectra. We hypothesize that since water lines cover a larger range of IGRINS' wavelength range compared to CO, the 1D retrieved Doppler shift is driven by the water lines, leading to a misalignment of the 1D spectra to the CO line cores. This moves the CO lines in the 1D spectra towards the line wings, which are shallower, where 1D spectra with lower chemical abundances provide a better match. This provides another explanation (alongside limitations of our P-T parameterization and degeneracies between the thermal and chemical structure, see Section~\ref{sec:discussion_p-t}) for why the CO lines in the 1D retrieved spectrum are shallower than the CO lines in the 3D phase-dependent GCM spectra.
Conversely, we note a mitigating factor to the Doppler offset between CO and H$_2$O may have been the spectral broadening due to winds and rotation. The rotational broadening is larger than the velocity offset between the H$_2$O and CO lines. This means the broadened CO lines in the 1D retrieval still have some overlap with the CO lines in the 3D GCM spectra, even when the retrieved orbital solution aligns the 1D spectra with the H$_2$O lines. This is advantageous if the goal is to constrain chemical abundances and the P-T structure, as it implies our 1D retrieval is less sensitive to velocity offsets between atmospheric species than in the case without broadened spectral lines.
\subsection{Caveats and limitations}
\label{sec:discussion_caveats}
Here we highlight key caveats and limitations associated with assumptions and choices in this study.
\begin{itemize}
\item \textit{Observational framework simulating HRS}: We opted for a simplified observational setup. This was done as the main goal of the paper is to investigate the impact of running a 1D retrieval on inherently 3D HRS data. Certain observational effects like telluric lines need to be included in order to enable post-processing with PCA, and subsequent model reprocessing within the 1D retrieval. These post-processing steps are not optional, as they remove the continuum information, which impacts the retrieved results as discussed in the previous sections. On the other hand, when adding too much realism in terms of observational effects, such as including airmass, frame-dependent wavelength solutions, time-varying telluric lines, blaze functions, this work would be investigating the impact of observational nuances on the retrieved results \citep[for a recent study, see][]{Savel2024} rather than 3D atmospheric effects. This motivated our choice for a constant continuum $S/N$ and stationary telluric lines.
\item \textit{GCM of WASP-76 b}: We note the isothermal/non-inverted profiles inside the hot spot are somewhat specific to the radiative transfer routine used in this GCM. Since our GCM uses a "picket fence" scheme, it produces more physically realistic temperature profiles than the simpler ``double gray" approach, but still, this can result in more isothermal profiles than a more multi-wavelength routine would produce \citep{Parmentier2015}. Yet, this GCM provides an excellent case study of a dataset where the brightest regions do not correspond with those producing the strongest emission features.
\item \textit{Choice of P-T parameterization}: we limited our investigation to the P-T profile parameterization by \citet{Madhusudhan2009}. We chose this parameterization for its general flexibility, enabling a range of non-inverted and inverted thermal structures. However, our retrieved results do show some P-T parameters are degenerate with the chemical abundances, and it cannot sufficiently capture the different thermal gradients in the CO and H$_2$O line forming regions. Additionally, a different parameterization may better capture the deeper non-inverted parts of the atmosphere. Examples could include P-T profile parameterizations that essentially fit a continuous curve along a set (P,T)-points, e.g., parameterizations presented in \citet{Bazinet2024, Smith2024b}, may be advantageous in capturing this nuance.
\item \textit{Scale parameter}: A scale parameter, introduced by \citet{Brogi2019}, has been used to trace changes of the spectral line strength during the orbit \citep{vanSluijs2023, Herman2022, Pino2022}. Previous retrievals often include a scale parameter $a$ as a nuisance parameter \citep[e.g.][]{Brogi2019, Line2021, Gibson2022}. This is helpful if the exact planet-to-star scaling is unknown beforehand, as for actual observations. In this study, we fixed $a=1$ because the correct scaling for the planet-to-star flux is known, and the scale parameter also influences the line contrast, thereby impacting the retrieved thermal and chemical structures in a nontrivial manner. In adverse scenarios, the inclusion of the scale parameter may bias the retrieved results. Conversely, it may mitigate the regional variation of the spectrum by shrinking or stretching spectral features, in line with findings of low-resolution spectroscopy retrievals by \citet{Taylor2020}. Moreover, future inclusion of a phase-dependent scale parameter may account for some of the line-contrast variation as a function of phase seen in Figure~\ref{fig:dopon_spectra_comparison}.

\item \textit{Additional opacity sources}: We limited this experiment to CO and H$_2$O. As our results demonstrate, the interplay between the thermal and chemical profiles is complex, even when just two opacity sources are included. For comparison against real observations, we recommend including additional opacity sources such as OH and H$^-$. We anticipate that additional opacity sources will broadly increase the atmospheric opacity, muting spectral features, obscuring deeper parts of the atmosphere. How these sources will impact the retrieved thermal and chemical profiles will likely depend on what opacity source is driving the retrieval results, their probed pressure range, and velocity offsets.
\end{itemize}
\section{Conclusion} \label{sec:conclusion}
In this work, we investigated how intricate details of a 3D atmosphere affect the robustness of a 1D retrieval. This was done by simulating high-resolution spectra within an observational framework from 3D phase-dependent GCM spectra. These mock observations were subsequently post-processed following prevailing HRS methods. The results of the 1D retrieval were presented and contextualized by comparison to their 3D GCM input models.
The primary conclusion of this work is that a 1D retrieval on inherently 3D data can broadly constrain P-T profiles and chemical profiles within the range of spatial profiles and ratios present in the data, thus recovering atmospheric conditions present at least in part of the atmosphere. Nevertheless, 1D retrieval results are not robust; the retrieved P-T and chemical profiles are not a homogeneous average of all spatial and phase-dependent information. HRS is more sensitive to spatial regions with large thermal gradients around pressures where the spectral lines form. These do not necessarily coincide with the planet's region emitting the most radiation, resulting in a bias in retrieved atmospheric properties.
While this was concluded from a limited number of experiments, for a specific 3D GCM, observational simulator, and retrieval parameterization, this provides an important first step towards the complexity of characterizing exoplanet atmospheres using HRS emission retrievals. Challenges include incorporating phase-dependence within a retrieval framework, allowing for molecule-dependent velocity offsets, and exploring a wider range of thermal and chemical profile parameterizations. To address these challenges, we will suggest directions for future research. Firstly, the impact of combining both low- and HRS can be considered, as previous studies have shown these methods are highly complementary \citep[e.g.][]{Brogi2018, Brogi2019, Smith2024}. Similar future studies may consider fitting for at least two P-T profiles, one to capture the absolute flux and one to capture the relative flux or explore the impact of explicitly including a (possibly phase-dependent) scale parameter. Nevertheless, a single P-T profile may still be an accurate approximation when those two regions largely overlap, for example, when the hottest regions also have the strongest temperature gradients, where line pressures form. Secondly, a study investigating the impact of different P-T and chemical profile parameterizations, and their impact on retrieved results, is warranted. This was demonstrated by: (1) the inherent degeneracies between the retrieved thermal profile and the chemical abundances when using the \citet{Madhusudhan2009} parameterization, and (2) its lack of flexibility to adjust to the different thermal gradients in the CO and H$_2$O line-forming layers. Thirdly, a similar study can be conducted but focusing on different 3D models for different exoplanets to investigate the impact of atmospheric complexity on 1D retrievals. Lastly, future studies could use 3D models and simulated data to focus on a more detailed analysis of how the velocity signatures impact retrieval results. Atmospheric species of interest are e.g. CO, H$_2$O, OH, and iron, where some recent studies have made GCM predictions \citep{Beltz2024, Wardenier2025}. We found that line broadening due to rotation and winds induced by atmospheric motion means that one may be less sensitive to small offsets between lines in the 1D retrieval forward model and the 3D GCM spectra, somewhat mitigating the effects of minor velocity offsets between species and or orbital phases on 1D retrieval results than in the absence of any line broadening. However, the problem of not including opacity source-dependent velocity offsets within a retrieval persists and may explain the bias towards lower chemical abundances observed between our 1D retrieved chemical abundances and those present in the 3D atmosphere. We expect this issue to be particularly important at higher instrumental spectral resolution, or when investigating slower-rotating exoplanets.
\software{The software accompanying this work is available on the first author's GitHub\footnote{\url{https://github.com/lennartvansluijs/1D_Retrievals_vs_3D_GCM}} and archived in Zenodo \citep{lennart_van_sluijs_2025_15978590}, Astropy \citep{Astropy2013, astropy:2018, astropy:2022},
Matplotlib \citep{Matplotlib2007}, Numpy \citep{Harris2020}, Scipy \citep{Virtanen2020}, {\sc Pymultinest} \citep{Feroz2008, Feroz2009, Feroz2019}}
\begin{acknowledgments}
We would like to thank the anonymous reviewers for their insightful comments and constructive feedback, which have significantly improved the quality of this paper. JLB and LvS acknowledge funding from the European Union's Horizon 2020 research and innovation program under grant agreement No 805445. This work also received financial support from the Heising-Simons Foundation, Grant \#2019-1403. A portion of this research was carried out at the Jet Propulsion Laboratory, California Institute of Technology, under a contract with the National Aeronautics and Space Administration (80NM0018D0004).
\end{acknowledgments}
\bibliography{references}{}
\bibliographystyle{aasjournal}
\appendix
\section{1D mock retrieval}
\label{sec:mock_retrieval}
\begin{figure}[ht]
    \centering
    \includegraphics[width=0.95 \textwidth]{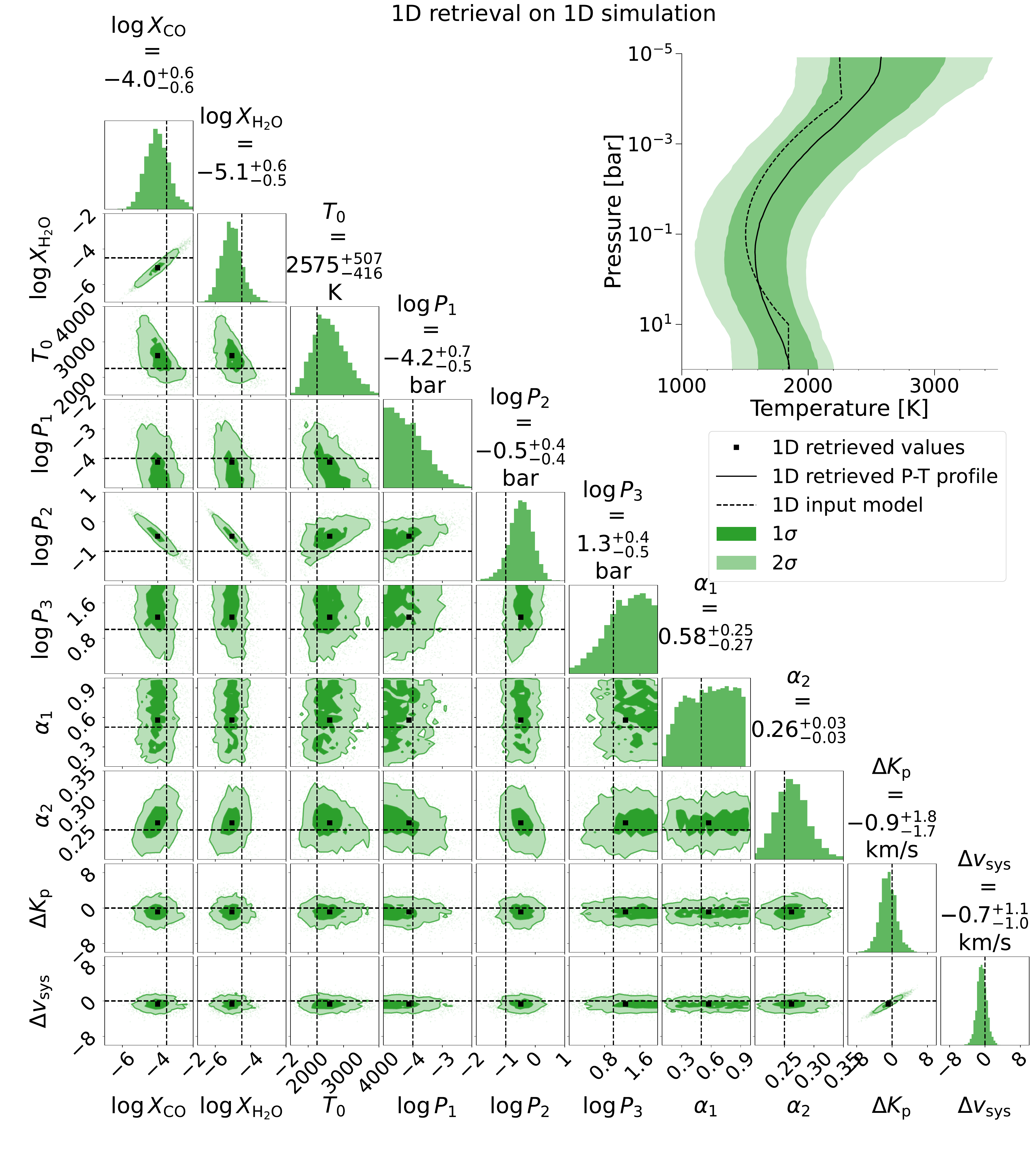}
    \caption{Marginalized distributions (corner plot) from the 1D retrieval on the 1D simulated data set to verify the retrieval algorithm can retrieve the correct 1D input parameters. Confidence intervals corresponding with 1$\sigma$ and 2$\sigma$ are shown in dark green and light green, respectively. The dashed black line shows the 1D forward model input parameters and P-T profile. 1D retrieved parameters are indicated by the black squares. \textit{Top-right:} the 1D retrieved P-T profile (solid black line), the 1$\sigma$ and 2$\sigma$ confidence intervals, and the input P-T profile (dashed black line).}
    \label{fig:corner_1D}
\end{figure}

We ran a mock retrieval to verify the results of our 1D retrieval framework. First, we created simulated HRS data following the recipe of Section~\ref{sec:sims}, but replacing the 3D GCM phase-dependent spectra with phase-independent 1D spectra. To compute the 1D spectra, we used the same forward model inherent to the retrieval framework. The input parameters are listed in Table~\ref{table:1d_sim_params}. They were chosen such that the 1D profiles are broadly within the range of the P-T and chemical profiles in the GCM of WASP-76 b. We follow the retrieval methodology described in Section~\ref{sec:retrieval}. Since the input model was unbroadened, no Doppler-broadening parameter was included in this retrieval.
The mock retrieval's parameter constraints are listed in Table~\ref{table:1d_sim_params} and the marginal distributions (corner plots) are shown in Fig~\ref{fig:corner_1D}. These demonstrate all parameters are within 1$\sigma$ of the input values, except for $\log P_2$, which is within 2$\sigma$. The relatively large uncertainties in some parameters defining the P-T profile, such as $\log P_2$, $\alpha_1$, and $T_0$, are the result of inherent degeneracies in the \citet{Madhusudhan2009} P-T parameterization. The key point is that the retrieved 1$\sigma$ confidence interval captures the 1D input P-T profile as shown in the right-top panel of Figure~\ref{fig:corner_1D}, and chemical abundances.
\begin{deluxetable}{l c r}[ht]
\tablecaption{1D simulation forward model input parameters. \label{table:1d_sim_params}}
\tablewidth{0pt}
\tablehead{
\colhead{\textbf{Parameter}} & \colhead{\textbf{Input value}} & \colhead{\textbf{Retrieved value}}
}
\startdata
\textbf{Chemical abundances} \\ \hline
\quad $\log(X_{\mathrm{CO}})$ & -3.5 & $\rm{-4.0^{+0.6}_{-0.6}}$ \\ 
\quad $\log(X_{\mathrm{H_2O}})$ & -4.5 & $\rm{-5.1^{+0.6}_{-0.5}}$ \\ \hline
\textbf{P-T profile} \\ \hline
\quad $T|_{P = 1\, \text{\rm{$\mu$bar}}}$ (K) & 2250 & $\rm{2575^{+507}_{-416}}$ \\ 
\quad $\log(P_1 \, (\mathrm{bar}))$ & -4 & $\rm{-4.2^{+0.7}_{-0.5}}$ \\ 
\quad $\log(P_2 \, (\mathrm{bar}))$ & -1 & $\rm{-0.5^{+0.4}_{-0.4}}$ \\ 
\quad $\log(P_3 \, (\mathrm{bar}))$ & 1 & $\rm{1.3^{+0.4}_{-0.5}}$ \\ 
\quad $\alpha_1 \, (\mathrm{K^{-1}})$ & 0.5 & $\rm{0.58^{+0.25}_{-0.27}}$ \\ 
\quad $\alpha_2 \, (\mathrm{K^{-1}})$ & 0.25 & $\rm{0.26^{+0.03}_{-0.03}}$ \\ \hline
\textbf{HRS Parameters} \\ \hline
\quad $\Delta K_{\rm{p}} \, \mathrm{km \, s^{-1}}$ & 0 & $\rm{-0.9^{+1.8}_{-1.7}}$ \\ 
\quad $\Delta V_{\mathrm{sys}} \, (\mathrm{km \, s^{-1}})$ &  0 & $\rm{-0.7^{+1.1}_{-1.0}}$ \\ 
\enddata
\end{deluxetable}
\section{Marginal distributions: 1D retrieval on 3D GCM simulation}
\label{sec:corner_plots}
\begin{figure}[h!]
    \centering
    \includegraphics[width=0.99 \textwidth]{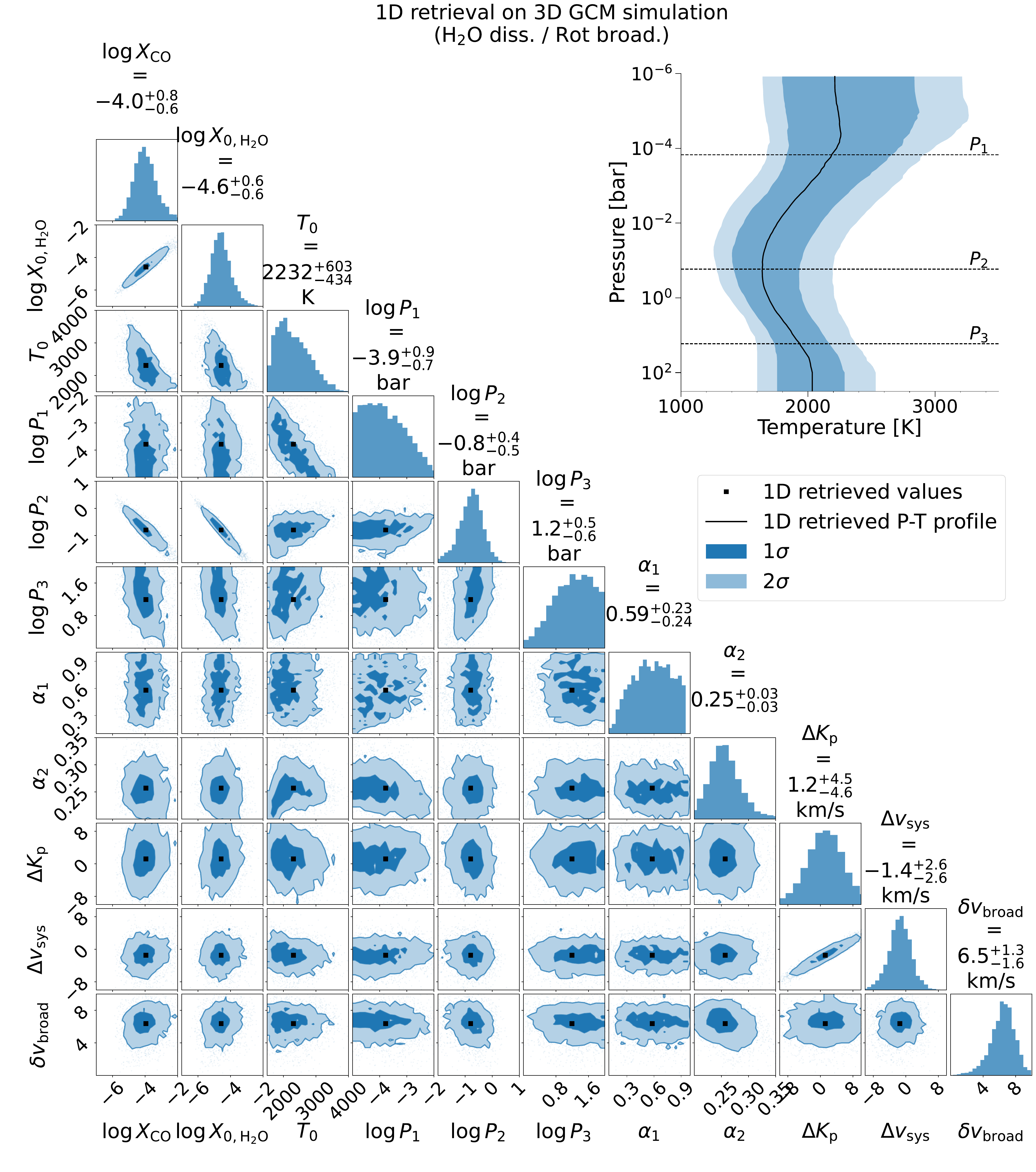}
    \caption{Marginalized distributions (corner plot) from the 1D retrieval on the 3D GCM simulated HRS data set show that part of the explored multi-dimensional parameter space is significantly better matched to the data. Confidence intervals corresponding with 1$\sigma$ and 2$\sigma$ are shown in dark blue and light blue respectively. Retrieved parameters are indicated by the black squares.  In this retrieval run, water dissociation was included and thus the retrieved water abundance corresponds to the bulk water abundance at the deepest pressure level, namely $X_{\text{0,H$_2$O}}$. \textit{Top-right panel:} shows the retrieved 1D P-T profile in black. The pressure levels corresponding with $P_1$, $P_2$, and $P_3$ are highlighted by the dashed lines.}
    \label{fig:corner_3D_dopoff_H2O_dissociated}
\end{figure}

For completeness, Figure~\ref{fig:corner_3D_dopoff_H2O_dissociated} shows the marginal distributions for the retrieval that included water dissociation and rotational broadening. The marginal distributions show that part of the explored multi-dimensional parameter space is significantly better matched to the data. Correlations between the abundance of H$_2$O and CO, and $v_{\rm{sys}}$ and $K_{\rm{p}}$ as seen in other retrievals with well-constrained parameters \citep[e.g.][]{Brogi2019}. The correlation between $\log{P_2}$ and the chemical abundances of water and CO abundances is expected because $\log{P_2}$ essentially sets the continuum level of emission features in this P-T parameterization. To match the spectral line shape in the data, one can either have a deeper continuum level with a lower chemical abundance, or a less deep continuum level with a higher chemical abundance.

\end{document}